\documentclass[10pt,pra,letterpaper,typeset,superscriptaddress,twocolumn,showpacs, floatfix]{revtex4-2}

\usepackage{amsmath}
\usepackage{amsfonts}
\usepackage{amssymb}
\usepackage{xcolor}
\usepackage{graphicx}
\usepackage{float}
\usepackage{xcolor}
\usepackage{hyperref}
\usepackage{verbatim}
\usepackage[english]{babel}
\immediate\write18{texcount -tex -sum  \jobname.tex > \jobname.wordcount.tex}

\providecommand{\keywords}[1]
{
  \small	
  \textbf{\textit{Keywords---}} #1
}

\begin{document}

\title{Preparing pre-established quantum correlations with pulse-level control}

%\title{Pulse-level resource control for trainable quantum circuits}
%\title{Beyond quantum gates: direct state preparation through optimized pulses}

%\title{Characterizing Entanglement Classes via Pulse-Based Quantum Control in Multipartite Systems}

\author{K. De La Ossa Doria}
\affiliation{Departamento de F{\'i}sica, Facultad de Ciencias B{\'a}sicas,
Universidad de Antofagasta, Casilla 170, Antofagasta, Chile}

\author{T. Merlo Vergara}
\affiliation{Instituto de F\'isica, Pontificia Universidad Cat\'olica de Chile, Casilla 306, Santiago, Chile}

\author{D. Goyeneche}
\affiliation{Instituto de F\'isica, Pontificia Universidad Cat\'olica de Chile, Casilla 306, Santiago, Chile}

\date{\today} % Comment this line to show today's date

\begin{abstract}
 In this work, we present a technique for quantum state preparation in transmon qubit system that employs optimized electromagnetic pulse sequences rather than discrete quantum gates. By focusing directly on quantum correlations rather than specific target states, we optimize the pulse parameters to generate the desired entanglement structures in two and three qubits systems. The resulting protocols reproduce the characteristic entanglement properties associated with the Bell, GHZ, and W classes, reaching values of the corresponding entanglement measures consistent with these entanglement structures.  The approach uses a reduced set of parameters to characterize the control pulses, thereby restricting the space explored during optimization while retaining the entanglement properties of interest.

% Minimizing the time required for quantum state preparation is crucial to mitigate decoherence and enable practical quantum algorithms on near-term hardware. In this work, we introduce a technique for quantum state preparation in transmon-qubit systems using optimized electromagnetic pulse sequences rather than discrete quantum gates. By directly targeting quantum correlations instead of specific target states, we identify minimal-time pulse protocols that optimize relevant entanglement resources, such as concurrence and the three-tangle for two and three qubit systems, respectively. For the figures of merit considered, this approach successfully achieves maximal entanglement in each case: Bell, GHZ and W like states. Beyond state preparation, the resource-oriented nature of the approach leads to a reduced effective expressivity of the control scheme, a feature that represents an advantage in algorithmic settings where excessive control freedom is known to hinder performance.
\end{abstract}


\maketitle

%TC:ignore
 
\section{Introduction}
Quantum computing has emerged as one of the most promising paradigms for next-generation information processing technologies. Despite rapid theoretical and experimental advances, the scalable realization of quantum processors remains limited by decoherence, control imperfections, and limited gate fidelities. To partially overcome these limitations, significant efforts have focused on advanced quantum control techniques aimed at optimizing system dynamics and enabling the precise implementation of quantum states and gates. In particular, quantum optimal control provides a general framework for designing sequences of electromagnetic pulses that maximize the fidelity with respect to target unitary transformations \cite{chakrabarti2004quantum,grivopoulos2005optimal,koch2022quantum}. This approach has been successfully applied to effective models describing a wide range of physical platforms, including transmon-type superconducting qubits \cite{krantz2019guide}, trapped ions \cite{ringbauer2022universal}, spins in quantum dots \cite{castro2022optimal,zhang2025operation}, and neutral atoms \cite{henriet2020quantum}, among others.

Among the first algorithms developed in this field are Krotov \cite{wilhelm2020introduction,schirmer2011efficient,abracontrol} and GRAPE \cite{ansel2024introduction,malarchick2025verified}, both based on gradient-optimization methods that iteratively refine the control pulses. More flexible approaches such as GOAT \cite{machnes2018tunable} enable direct optimization over analytic families of control signals. More recently, machine learning techniques have been incorporated into control design, including deep reinforcement learning \cite{wang2022quantum} and hybrid classical–learning-based strategies \cite{genois2025quantum}, expanding the range of physically realizable control protocols. Quantum optimal control plays a central role in physics and chemistry, where precise manipulation of quantum dynamics is required to reach target states, implement unitary operations, or achieve controlled population transfer between specific energy eigenstates \cite{georgefrancis2025data,nguyen2024reinforcement}.

Beyond gate synthesis and state preparation, quantum control is increasingly relevant for the practical implementation of variational quantum algorithms on noisy intermediate-scale quantum (NISQ) devices \cite{wang2024pulse}. Variational approaches rely on parametrized quantum circuits optimized through classical feedback, and their performance is strongly influenced by the expressivity of the chosen parametrization \cite{holmes2022connecting}. 
%While highly expressive circuits can represent a broad class of quantum states, growing theoretical and numerical evidence indicates that excessive expressivity can induce barren plateau phenomena, where cost-function gradients become exponentially suppressed with system size, rendering optimization impractical \cite{larocca2025barren}. This trade-off between expressivity and trainability highlights the need for control strategies that restrict the accessible dynamical space in a physically informed manner.

From this perspective, pulse-level quantum control provides a natural alternative to standard gate-based constructions. By operating directly on the time-dependent Hamiltonian of the system, pulse-based approaches inherently constrain the effective expressivity in the evolution to what can be achieved within a given time window and hardware configuration. Rather that being a drawback, this controlled reduction in expressivity can suppress error accumulation, simplify optimization landscapes, and improve robustness and trainability in variational settings, while still enabling the generation of relevant quantum resources.

Recent studies have explored quantum optimal control as a means to manipulate entanglement directly, without relying on fixed gate decompositions or predefined target states. Romano and Del Fabbro \cite{del2007optimal} analyzed the optimal generation of bipartite entanglement under local control, while Platzer, Mintert, and Buchleitner \cite{platzer2010optimal} and Lucas \emph{et al.} 
\cite{lucas2013tailoring} formulated dynamical-control frameworks in which multipartite entanglement measures enter explicitly as cost functionals. Complementary pulse-based and variational approaches have demonstrated efficient entanglement generation under realistic constraints in superconducting and hybrid architectures \cite{del2007optimal}, providing a solid foundation for quantum resource oriented pulse optimization.

In this work, we focus on identifying the simplest possible set of electromagnetic pulse protocols that generate a prescribed set of quantum correlations. By emphasizing the preparation of quantum resources rather than reproducing predefined quantum states, our approach substantially reduces the control cost required to achieve a given level of entanglement. We illustrate the method by considering minimal two- and three-qubit systems and designing pulse sequences that directly optimize resource-oriented objectives: \emph{(a)} for two qubits, the maximization of quantum negativity; and for three qubits, \emph{(b)} the maximization of the 3-tangle, and \emph{(c)} the minimization of the cost function $\sqrt{\sum_{i<j}\left( C_{ij} - \frac{2}{3}\right)^{2}}$, $i,j=1,2,3$, where $C_{ij}$ denotes the concurrence of the bipartite reduction $\rho_{ij}$ \cite{Hill1997entanglement}. For the open-system case, we use Genuine Multipartite Negativity \cite{jungnitsch2011taming,hofmann2014analytical} hese representative cases define fundamental control primitives that can be systematically extended to construct less expressive, resource-efficient pulse-based implementations of more general quantum circuits.

\section{Superconducting Qubits}\label{superconducting} 

Superconducting circuits constitute one of the most versatile platforms for the physical realization of quantum bits. They enable precise control of quantum states through microwave pulses applied to nonlinear resonant elements operating at cryogenic temperatures \cite{wei2023compact,kang2025new}. Among the different implementations, the transmon qubit has become a standard architecture due to its long coherence time, high-fidelity gate operations, and a simple integration into scalable multi-qubit systems \cite{koch2007charge}. These systems can be accurately modeled as weakly anharmonic oscillators, where quantum information is encoded in the two lowest energy levels of the circuit.

The dynamics of an $n$-qubit superconducting processor driven by external control fields is governed by a time-dependent Hamiltonian that can be expressed as
\begin{equation}
    H(t) = H_{\text{sys}} + \sum_{j=1}^{n} \mu_{i}(t) H_{j},
    \label{eq_1}
\end{equation}
where $H_{\text{sys}}$ represents the intrinsic (drift) Hamiltonian of the system, and the second term corresponds to the set of control Hamiltonians $H_{j}$weighted by time-dependent amplitudes $\mu_{j}(t)$. These control fields are typically implemented as shaped microwave pulses that induce transitions between energy levels or generate effective couplings between qubits. The objective of optimal control is to determine the set of functions $\{\mu_{i}(t)\}$ and their parameters that drive the system from an initial state $|\psi_{0}\rangle$ to a target state $|\psi_{\mathrm{t}}\rangle$, while maximizing a cost functional $\mathcal{F}[\mu_{j}(t)]$ associated with fidelity, entanglement, or other physical figure of merit.

The explicit form of the Hamiltonian depends on the system dimensionality and whether the analysis addresses single-qubit dynamics or coupled multi-qubit configurations. In what follows, we present the effective Hamiltonians for one, two, and three transmon qubits, focusing on the introduction of control terms that enable the implementation of local operations and entangling interactions.

\subsubsection{Single Qubit}

The Hamiltonian of a single superconducting qubit can be written as \cite{krantz2019guide}:
\begin{equation}
    H_{1,\text{sys}} = \omega b^{\dagger} b + \frac{\alpha}{2}\, b^{\dagger} b^{\dagger} b b,
    \label{eq_2}
\end{equation}
where $\omega$ denotes the transition frequency between the two lowest energy levels $|0\rangle \leftrightarrow |1\rangle$, $\alpha$ is the anharmonicity, and $b$ ($b^{\dagger}$) is the annihilation (creation) operator of the transmon mode. Here,and throughout this work, we assume that $\hbar = 1$. The weak anharmonicity allows the transition between the first two energy levels to be distinguished from higher-level transitions, enabling the system to operate effectively as a qubit while retaining an accurate description of higher energy levels when necessary.

The transmon state is manipulated through an external electromagnetic drive pulse with frequency $\omega_{d}$, whose interaction with the circuit can be modeled as
\begin{equation}
    H_{1,\text{ctl}}(t) = \Omega_{d}\, \mathrm{Re}\!\left[\mu(t)\, e^{i \omega_{d} t}\, (b^{\dagger} + b)\right],
    \label{eq_3}
\end{equation}
where $\Omega_{d}$ represents the drive strength and $\mu(t)$ is a dimensionless control envelope normalized to the interval $\mu(t) \in [-1,1]$. The drive Hamiltonian describes the coherent coupling between the qubit and the microwave field, enabling state rotations on the Bloch sphere. .

For a simplified mathematical description of the problem it is convenient to move into a \emph{rotating reference frame} at the drive frequency $\omega_{d}$ and apply the rotating-wave approximation (RWA) \cite{krantz2019guide} \cite{georgefrancis2025data,nguyen2024reinforcement}. This transformation removes fast-oscillating terms and leads to an effective Hamiltonian expressed as
\begin{equation}
    H_{\mathrm{RWA}} = \frac{\Delta}{2}\, \sigma_{z} + \frac{\Omega_{d}}{2}\, 
    \left[\mathrm{Re}\{\mu(t)\}\, \sigma_{x} + \mathrm{Im}\{\mu(t)\}\, \sigma_{y}\right],
    \label{eq_4}
\end{equation}
where $\Delta = \omega - \omega_{d}$ is the detuning between the qubit and drive frequencies, and $\sigma_{x}$, $\sigma_{y}$, $\sigma_{z}$ are the Pauli matrices. This effective Hamiltonian captures the essential dynamics of single-qubit control, where the envelope $\mu(t)$ directly determines the rotation axis on the Bloch sphere.

% Para generar correlaciones entre qubits, se emplea como compuerta nativa la llamada ECR (Echoed Cross Resonance), que aprovecha el acoplamiento entre transmons a través de resonadores para inducir interacciones efectivas de los dos qubits.  

% Podemos reescribir el Hamiltoniano mediante la siguiente transformación $R(t) = e^{-i\omega_{d}tb^{\dagger}}b$ y usando la aproximación de onda rotatoria (RWA), llegamos al siguiente Hamiltoniano.

% \begin{multline}
%     H_{1}(t ) = R^{\dagger}(t)\left(H_{1,sys} + H_{1,ctl}(t) - i\partial_{t}  \right)R(t) \\ \approx \delta b^{\dagger}b + \frac{\alpha}{2}b^{\dagger}b^{\dagger}bb + \left[\frac{\Omega_{d}}{2}\mu(t)b + h.c \right]
% \end{multline}

% Aquí $\delta =\omega - \omega_{d}$ es el detuning, el cual desaparece cuando las frecuencias entran en resonancias , i.e.m $\omega = \omega_{d}$, Ahora bien si consideramos los dos primeros niveles de energía y remplazamos $b \rightarrow \sigma{-} = X - iY$ y $b^{\dagger}\rightarrow \sigma^{\dagger}=X+iY$ en el Hamiltoniano de control, se obtiene como resultado

% \begin{equation}
%     H_{1,ctl}(t) = \frac{\Omega_{d}}{2}\textbf{Re}(\mu(t))X + \frac{\Omega_{d}}{2}\textbf{Im}(\mu(t))Y 
% \end{equation}

\subsubsection{Two qubits}
\label{sub: two qubits}
In this configuration, we consider a system composed of two capacitively coupled transmon qubits, each characterized by intrinsic physical parameters such as the transition frequency and the anharmonicity. The energy levels of each qubit can be individually manipulated by applying microwave pulses through independent control channels. When the drive frequency is tuned in resonance with the desired transition frequency of a given qubit, and the pulse amplitude, phase, and temporal shape are appropriately modulated, precise local gates can be implemented with high fidelity \cite{sundaresan2020reducing}. 

The implementation of two-qubit operations relies on dynamically modulating the interaction between the transmons. This can be achieved by applying an additional off-resonant drive to one of the qubits, whose frequency is chosen to activate a specific coupling channel with the target qubit. Under this driving scheme, the effective Hamiltonian describing the coupled system is given by
\begin{equation}
\begin{split}
    H_{2,\text{sys}} &= \sum_{j=0}^{1} \left(\omega_{j} b_{j}^{\dagger} b_{j} + \frac{\alpha_{j}}{2}\, b_{j}^{\dagger} b_{i}^{\dagger} b_{j} b_{j}\right) \\ 
    &+ J_{01}\left(b_{0} b_{1}^{\dagger} + b_{0}^{\dagger} b_{1}\right),
    \label{eq_5}
\end{split}    
\end{equation}
where $\omega_{j}$ denotes the transition frequncy between the first two levels of the $i$th qubit, $\alpha_{j}$ is its anharmonicity, and $J_{01}$ is the coupling strength between the two transmons. The operators $b_{j}$ ($b_{j}^{\dagger}$) are the annihilation (creation) operators acting on the corresponding local Hilbert space.

The control Hamiltonian associated with the external drives can be expressed as
\begin{equation}
\begin{split}
    H_{2,\text{ctrl}}(t) = \sum_{j=0}^{1} \Omega_{d_{j}}\,
    \mathrm{Re}\!\left[\mu_{j}(t)\, e^{-i(\omega_{d_{j}} t + \phi_{j})}\right] 
    (b_{j}^{\dagger} + b_{j}),
    \label{eq_6}
\end{split}
\end{equation}
where $\Omega_{d_{j}}$ is the drive amplitude for the $j$th qubit, $\omega_{d_{j}}$ the drive frequency, $\phi_{j}$ is the drive phase, and $\mu_{j}(t)$ is the normalized control envelope. By tuning these parameters, both local and effective nonlocal interactions can be engineered, enabling the implementation of high-fidelity entangling gates such as  controlled-phase or iSWAP gates, depending on the specific detuning and pulse conditions.

\subsubsection{Three qubits}
We now extend the model to a system of three transmon qubits, labeled from right to left as $0$, $1$, and $2$. In this configuration, qubit 0 acts as a control for qubit 1, while qubit 1 controls qubit 2, resulting in a linear, unidirectional coupling topology. Although the qubits are physically connected, bidirectional control is not assumed. The system Hamiltonian is given by
\begin{multline}
     H_{3,\text{sys}} = 
     \sum_{j=0}^{2} \left(\omega_{j} b_{j}^{\dagger} b_{j} + \frac{\alpha_{j}}{2}\, b_{j}^{\dagger} b_{j}^{\dagger} b_{j} b_{j}\right) \\
     + J_{01}\left(b_{0} b_{1}^{\dagger} + b_{0}^{\dagger} b_{1}\right) 
     + J_{12}\left(b_{1} b_{2}^{\dagger} + b_{1}^{\dagger} b_{2}\right),
     \label{eq_7}
\end{multline}
where $\omega_{j}$, $\alpha_{j}$, and $J_{jk}$ denote the transition frequency, anharmonicity, and coupling strength, respectively, for each pair of interacting qubits. The ladder operators $b_{j}$ ($b_{j}^{\dagger}$) act locally on the Hilbert space of qubit $j$.

As in the two-qubit case, the transmons are individually controlled by microwave pulses applied to each qubit through dedicated drive lines. The corresponding control Hamiltonian reads
\begin{multline}
    H_{3,\text{ctrl}}(t) = 
    \sum_{i=0}^{2}\Omega_{d_{j}}\, 
    \mathrm{Re}\!\left[\mu_{j}(t)\, e^{i \omega_{d_{j}} t}\right]
    (b_{j}^{\dagger} + b_{j}),
    \label{eq_8}
\end{multline}
where $\Omega_{d_{j}}$, $\mu_{j}(t)$, and $\omega_{d_{j}}$ represent the drive amplitude, control envelope, and drive frequency for each qubit, respectively. As before, local gates can be are implemented by applying resonant pulses at the corresponding transition frequencies through independent control channels, while for two-qubit gates the implementation consists of sending a pulse to one of the qubits (the control qubit) at the transition frequency of the qubit to be controlled (the target qubit).

The central challenge of this work consists of determining the optimal set of pulse parameters, including amplitudes and durations, that generates the characteristic entanglement structures associated with each inequivalent class of multipartite entanglement. The optimization is formulated to maximize the corresponding entanglement measures.

% como se puede notara las operaciones de control son realizadas localmente sobre los qubits, además las constante de acoplamiento se encuentran fijas, esto puede llevar que al querer controlar uno de los qubits de forma individual halla transición sobre uno de los otros qubits, tampoco podemos quitar de la dinámica la presencia o la existencia de los demás sistemas al momento querer controlarlos de manera individual, esto con lleva a que se produzca una especie de ruido.

\section{Quantum decoherence}

In previous sections, we discussed superconducting qubits described by system and control Hamiltonians. Each Hamiltonian represents a different physical contribution to the dynamics $H_{j,\mathrm{sys}}$ describes the superconducting circuit, while $H_{j,\mathrm{ctrl}}$ describes the electromagnetic pulses generated by arbitrary waveform generators (AWGs). In most case, the dynamics of these interacting are considered under the assumption of a \textit{closed system}. However, realistic quantum systems are never completely isolated from their environment and inevitably interact with it \cite{krantz2019guide}. 
\\

The interaction between a quantum system and its environment causes the system to become entangled with many environmental degrees of freedom. This interaction modifies the system's dynamics and affects the outcomes of measurements performed on it. As information about the system becomes correlated with the environment, quantum information effectively leaks from the system into inaccessible environmental degrees of freedom. This process is known as \textit{decoherence}. Decoherence can be viewed as a process that suppresses quantum coherences in the reduced state of the system, thereby favoring certain states, often referred to as pointer states, that are more robust against environmental interactions. This provides a physical mechanism underlying the emergence of classical behavior from quantum dynamics. Although the total system-environment evolves according to reversible quantum dynamics, the dispersal of information into inaccessible environmental correlations leads to effectively irreversible dynamics when only the reduced system is considered \cite{gautier2025optimal}. 

\subsection{Decoherence in superconducting qubits}

For a complete description of quantum dynamics, it is necessary to consider the interaction between the system and its environment. In quantum computers, mitigating decoherence remains a major challenge, as qubits inevitably interact with other qubits, their surrounding environment, and imperfections introduced during fabrication, all of which contribute to noise and degrade quantum coherence. The main decoherence mechanisms in superconducting qubits are \textit{relaxation} and \textit{dephasing}. These processes are commonly characterized by two experimentally measurable time scales, $T_{1}$ and $T_{2}$. The relaxation time $T_{1}$ characterizes the decay of the qubit population through energy exchange between the system and the environment. The coherence time $T_{2}$ characterizes the decay of the qubit's phase coherence and includes contributions from both energy relaxation and pure dephasing. The latter refers to the loss of phase coherence without energy exchange and can arise from temporal fluctuations in the qubit transition frequency  \cite{naeij2025open}.The relaxation and dephasing times can be related as \cite{kwon2021gate,krantz2019guide}

\begin{subequations} \label{eq:decohe}  \begin{align} \Gamma_{1} &= \frac{1}{T_{1}} \label{deco}
\\  \Gamma_{2} &= \frac{1}{T_{2}} = \Gamma_{\phi} + \frac{\Gamma_{1}}{2}   \label{depha}  \end{align} \end{subequations}

where $\Gamma_{1}$ is the population decay rate of the excited state and $\Gamma_{\phi}$ is the pure dephasing rate. In the relaxation process, spontaneous emission and thermal absorption are the two main mechanisms. If the qubit transition frequency is much larger than the thermal energy scale of the environment, i.e., $k_{B}T \ll \hbar\omega_{q}$, the contribution from thermal absorption can be neglected when calculating $\Gamma_{1}$.

\subsection{Lindblad master equation}

Whereas the dynamics of a closed quantum system can be described by unitary time evolution, the dynamics of an open quantum system cannot, in general, be described solely in terms of a unitary evolution. One of the most important approaches for modeling the dynamics of open quantum systems and their interaction with the environment is the master equation.
\\
In general terms, a quantum system S is open when it is coupled to another quantum system B, called the environment. The combined system is represented in Hilbert space as $\mathcal{H} = \mathcal{H}_{S}\otimes \mathcal{H}_{B}$ and the Hamiltonian takes the form 

\begin{equation}
    H(t) = H_{S}\otimes I_{B} + I_{S}\otimes H_{B} + H_{I},
    \label{eq_10}
\end{equation}

where $H_{S}$ is the Hamiltonian of the open system $S$, $H_{B}$ is the free Hamiltonian of the environment $B$, and $H_{I}(t)$ is the interaction Hamiltonian describing the coupling between the system and the environment. The dynamics of the subsystem $S$ resulting from the Hamiltonian evolution of the total system is commonly referred to as reduced system dynamics \cite{breuer2002theory}. To describe this dynamics, we make the following assumptions \cite{breuer2002theory,kwon2021gate}:

\begin{itemize}
    \item[1.]  Separability at $t=0$, there are no correlations between the system and its environment. Therefore, the total density matrix can be defined as a tensor product $\rho_{S}(0)\otimes\rho_{B}(0)$.
    
    \item[2]  Born approximation, the interaction between the system and environment is sufficiently weak. Moreover, the environment is much larger than the system, so changes in the density operator of the environment are negligible, and the system and the environment remain in an approximate product state at all times, i.e., 
    \begin{equation}
        \rho(t) \approx \rho_{S}\otimes\rho_{B}, 
        \label{eq_11}
    \end{equation}

    where $\rho_{B}$ is approximately constant at all times.

    \item[3] The Markov approximation; memory effects of the environment are negligible. In other words, the time scale of decay of environment correlation functions is much shorter than the time scale of the qubit dynamics. To obtain the dynamics of the qubit, we calculate the reduced density matrix of the qubit by performing a partial trace over the environmental degrees of freedom $(\rho_{S} = Tr_{B}\rho)$. In this way, the master equation governing the dynamics of the system can be obtained

     \begin{multline}
         \frac{d \rho_{S}(t)}{dt} = \\ -i\left[\hat{H}_{S}(t),\rho_{S}(t)\right] + \sum_{j=1}^{2}  \left( \hat{L}_{j} \rho_{S}(t) \hat{L}_{j}^{\dagger} - \frac{1}{2} \{ \hat{L}_{j}^{\dagger} \hat{L}_{j}, \rho_{S}\} \right)
         \label{lindblad}
     \end{multline}

%      \begin{multline}
% \frac{d \rho_{S}(t)}{dt} = 
% -i\left[\hat{H}_{S}(t),\rho_{S}(t)\right]
% +\sum_{i=1}^{2}
% \left(
% \hat{L}_{i}\rho_{S}(t)\hat{L}_{i}^{\dagger}
% -\frac{1}{2}
% \{
% \hat{L}_{i}^{\dagger}\hat{L}_{i},\rho_{S}(t)
% \}
% \right),
% \label{lindblad}
% \end{multline}

\end{itemize}

where $\{  \}$ shows the anticommutator, which is defined for the A and B operators as $\{A,B\} = AB + BA$, and $\hat{L}_{i}$ are the Lindblad (collapse) operators associated with the different decoherence channels.the corresponding decoherence rates are absorbed into the collapse operators and therefore do not appear as additional prefactors in Eq.~\eqref{lindblad}. For a single qubit, the relaxation and pure-dephasing processes are described by \cite{krantz2019guide} 

\begin{subequations}
\begin{align}
        \hat{L}_{1} &= \sqrt{\Gamma_{1}}\hat{\sigma}_{-}  \label{L1}\\
    \hat{L}_{2} &= \sqrt{\Gamma_{\phi}/2}\hat{\sigma_{Z}} \label{L2} 
\end{align}
\end{subequations}

where $\hat{L}_{1}$ describes relaxation, corresponding to the transition from $|1\rangle$ to $|0\rangle$, while $\hat{L}_{2}$ describes pure dephasing. The parameters $\Gamma_{1}$ and $\Gamma_{\phi}$ denote the relaxation and pure-dephasing rates, respectively. For superconducting qubits, these rates can be obtained from the relaxation and coherence times, $T_{1}$ and $T_{2}$, according to \cite{breuer2002theory}

\begin{equation}
\Gamma_{1}=\frac{1}{T_{1}},
\qquad
\Gamma_{\phi}=\frac{1}{T_{2}}-\frac{1}{2T_{1}}.
\label{rates}
\end{equation}
\\\\

Thus, the experimentally reported values of $T_{1}$ and $T_{2}$ determine the collapse operators used in the open-system simulations. Throughout this work, we use $\hbar=1$.
% Quantum computers based on superconducting qubits have the data for times $T_{1}$ and $T_{2}$, which is our last ingredient for making simulation of equation \ref{lindblad}

\section{Implementation} \label{imple}

\subsection{Three-qubit setup}

An optimal quantum control scheme was implemented to prepare tripartite entangled states—specifically the GHZ and $W$ classes—in a system of three superconducting transmon qubits. The physical parameters were extracted from the simulated backend \textit{Fake\_Sherbrooke} provided by IBM Quantum, which reproduces the Hamiltonian characteristics of the real \textit{ibm\_sherbrooke} device. This processor belongs to the \textit{Eagle} family of 127-qubit chips and features a hexagonal lattice connectivity, where each qubit is coupled to two or three neighbors. Such a layout reduces undesired crosstalk and supports high-fidelity native operations. The architecture natively implements a standard instruction set including ECR, ID, DELAY, MEASURE, RESET, RZ, SX, X, IF\_ELSE, FOR\_LOOP, and SWITCH\_CASE instructions \cite{abughanem2024ibm}.

\vspace{0.3cm}

The system dynamics were simulated using the \textit{Solver} module of \texttt{qiskit\_dynamics} (version 0.5.1), with the differentiable integrator \texttt{jax\_odeint}. This method provides efficient and accurate time evolution for systems governed by time-dependent Hamiltonians, even when multiple control channels are present. A key advantage of \texttt{jax\_odeint} is its adaptive-step integration scheme: unlike explicitly discretized methods, it automatically adjusts the time grid according to the local complexity of the evolution, allowing the capture of fast or strongly driven dynamics without manual specification of sampling points. Numerical precision and stability are ensured by user-defined absolute and relative tolerances.

\vspace{0.4cm}

The physical model considered consists of three linearly coupled transmon qubits. The qubits used were 7, 8, and 9, which we label as 0, 1, and 2, respectively. The Hamiltonian parameters are expressed as angular frequencies coefficients: $\omega_7\approx29.877\,\mathrm{rad/ns}$, $ \omega_8\approx30.235\,\mathrm{rad\,/ns}$, and $\omega_9\approx29.135\,\mathrm{rad\,/ns}$; anharmonicities $\alpha_7\approx-1.954\,\mathrm{rad\,/ns}$, $\alpha_8\approx-1.969\, \mathrm{rad\,/ns}$, and $\alpha_9\approx-1.839\,\mathrm{rad\,/ns}$; coupling strengths $J_{78}\approx0.013\,\mathrm{rad\,/ns}$ and $J_{89}\approx0.014\,\mathrm{rad\,/ns}$; and drive amplitudes $\Omega_{d0}\approx0.396\,\mathrm{rad\,/ns}$, $\Omega_{d1}\approx0.650\, \mathrm{rad\,/ns}$, and $\Omega_{d2}\approx0.315\,\mathrm{rad\,/ns}$. Whenever ordinary frequencies are quoted, they are obtained by dividing the corresponding angular-frequency coefficient by $2\pi$. In particular, for qubits 7 and 8, $|\omega_7-\omega_8|/(2\pi)=56.9437\,\mathrm{MHz}$ and $J_{78}/(2\pi)=2.013\,\mathrm{MHz}$.

\vspace{0.3cm}

Control fields were generated using the \textit{qiskit.pulse} module as piecewise-constant microwave drives applied through the channels $d0$, $d1$, $d2$, $U0$, and $U1$. Each pulse was characterized by two main parameters, its duration and amplitude, allowing the control sequence to be represented as a series of constant-amplitude segments. These signals define a set of time-dependent control functions $\mu_j(t)$ whose support extends over the total simulation time $T$.The simulation time was discretized in integer multiples of the backend's native time resolution, $dt = 0.2222$ ns, which sets the minimum temporal sampling interval for the control pulses.

\vspace{0.1cm}

The optimization of the control parameters was performed using the \textit{Differential Evolution} algorithm, a population-based global optimization method well suited to nonconvex objective landscapes, as commonly encountered in quantum optimal control. The algorithm performs a broad exploration of the parameter space, reducing the likelihood of premature convergence to local optima. The cost function $\mathcal{F}[\mu(t)]$ was defined in terms of the normalized final state of the system and formulated to quantify the characteristic entanglement properties associated with the desired multipartite entanglement class. Distinct entanglement-based metrics were employed for the GHZ and $W$ classes, allowing the optimization procedure to be adapted to their respective correlation structures. This formulation provides a unified numerical framework for designing pulse-based control protocols that reproduce the characteristic entanglement properties of genuinely tripartite states in superconducting architectures.

\subsection{Master equation implementation} 
\label{Master equation implementation}

For the implementation of Eq.~\ref{lindblad}, we consider systems of two and three qubits, where the system Hamiltonian is given by the sum of the individual qubit Hamiltonians, the coupling Hamiltonians between them, and the corresponding control terms. Furthermore, IBM quantum computers provide coherence-time data, $T_{1}$ and $T_{2}$;which we obtained from \textit{Fake\_Sherbrooke}. Each qubit has its own characteristic relaxation and coherence times, which are required to determine the corresponding decoherence rates. We use Eqs.~\ref{deco} and \ref{depha} to calculate $\Gamma_{1}$ and $\Gamma_{2}$, respectively, for each qubit.
\\\\
In this model, the decoherence times for qubits 7, 8, and 9 are $T_{1}^{(7)} \approx 230522.191$ ns and $T_{2}^{(7)} \approx 18177.314$ ns, $T_{1}^{(8)} \approx 394165.106$ ns and $T_{2}^{(8)} \approx 17282.471$ ns, and $T_{1}^{(9)} \approx 291846.552$ ns and $T_{2}^{(9)} \approx 56346.803$ ns, respectively.The Lindblad equation was implemented using \textit{Qiskit Dynamics} (version 0.5.1), where the \texttt{Solver} was configured with the \texttt{static\_dissipator} argument to incorporate the dissipative terms. The relaxation and dephasing processes are described by the dissipators given in Eqs.~\ref{L1} and \ref{L2}, respectively, where $\sigma_{-}$ denotes the lowering operator and $\sigma_{Z}$ denotes the Pauli-$Z$ operator.
\\
The pulse sequences considered in the open-system simulations were those previously obtained for the corresponding closed-system cases. To assess whether these control protocols remain effective in the presence of decoherence, we evaluated the resulting entanglement using the corresponding density matrices. For the two-qubit case, the negativity was used as the entanglement measure and was calculated from the final density matrix. For the three-qubit system, we restricted the analysis to the GHZ class and used the pulse sequence obtained from the closed-system optimization. No further optimization of the pulse parameters was performed in the presence of noise, allowing us to directly assess the robustness of the closed-system control protocol under dissipative dynamics.

To compare the results obtained from the closed- and open-system dynamics, the Genuine Multipartite Negativity (GMN) was used as the figure of merit, as it provides a measure of genuine tripartite entanglement for both pure and mixed states. This property is particularly relevant for open systems, where interactions with the environment can lead to mixed states as a result of decoherence. The GMN was evaluated using a computational implementation based on a semidefinite program \cite{jungnitsch2011taming,hofmann2014analytical}.Thus, the GMN provides a common criterion for evaluating and comparing the effectiveness of the control protocols obtained from closed- and open-system simulations in generating and preserving genuine multipartite quantum correlations.

\section{Results}

\subsection{Bipartite entanglement}

\begin{figure}[H]
    \centering
    \includegraphics[width=1\linewidth]
    {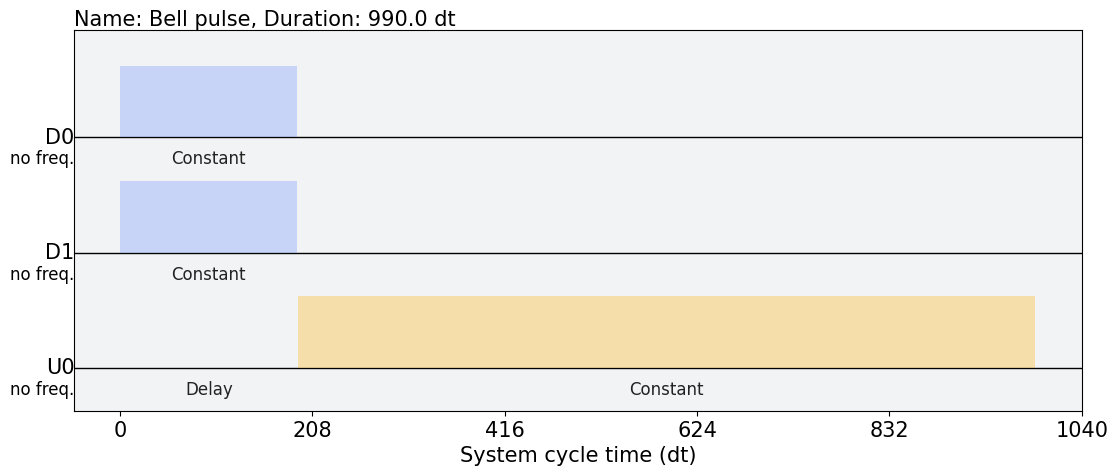}
    \caption{Pulse scheme used to prepare the maximally entangled state in Eq.~\ref{distri dos qubits}. The first stage consists of two single-qubit drive pulses (cyan) applied to qubits 7 and 8 through channels D0 and D1, respectively. Since the static coupling $J_{78}$ in Eq.~\ref{eq_5} remains active during these pulses, the total evolution is not strictly local and some entanglement may already develop in this interval. Afterwards, a cross-resonant drive (yellow) is applied through channel U0 by driving qubit 7 at the transition frequency of qubit 8, thereby enhancing the entangling dynamics. The total duration of the pulse sequence is $989.0\,dt$.}
    \label{fig: pulse_bell2_square}
\end{figure}

In this work, we consider a control scheme that begins with two simultaneous local drive pulses applied to qubits 7 and 8, with equal durations and independently optimized amplitudes. Although these control fields act locally, the static inter-qubit coupling $J_{78}$ in Eq.~\ref{eq_5} remains active throughout the sequence; therefore, the total evolution during this initial stage is not strictly local, and a finite amount of entanglement may already develop. A subsequent cross-resonant drive is applied to qubit 7 at the transition frequency of qubit 8, thereby enhancing the entangling dynamics between the two qubits, as shown in Fig.~\ref{fig: pulse_bell2_square}.

\begin{equation}
\begin{split} 
|\Psi\rangle  =     (-0.3684480152 + 0.5999791316i)|00\rangle \\
    -(0.0383474023 - 0.0330939344i)|01\rangle\\
    -(0.0167077292 + 0.0291402261i)|10\rangle \\
    +(0.2125136408 + 0.6748444424i)|11\rangle. 
    \end{split}
    \label{distri dos qubits}
\end{equation}

\begin{figure}[H]
    \centering
    \includegraphics[width=1\linewidth]
    {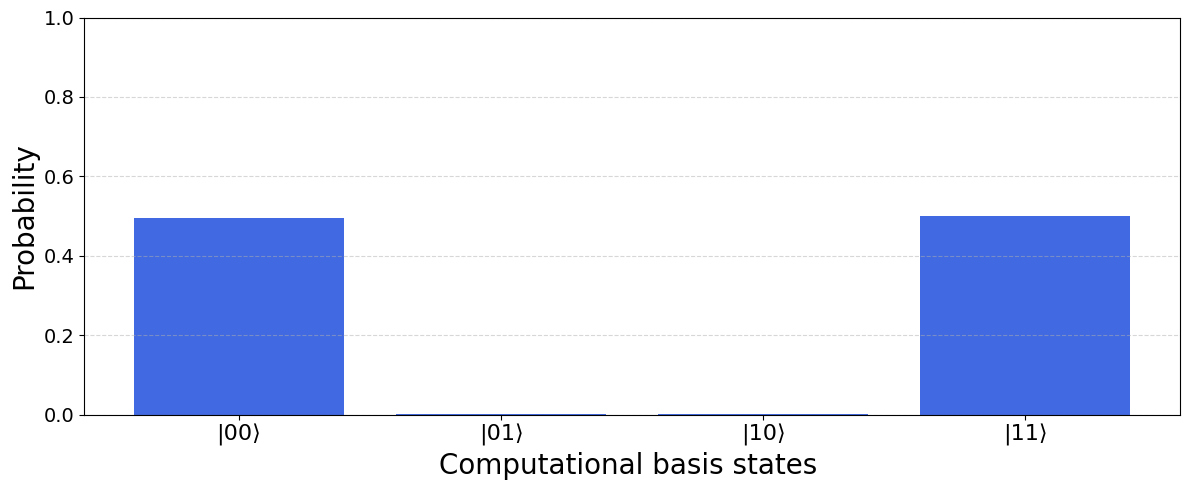}
    \caption{Probability distribution of the state in Eq.~\ref{distri dos qubits} in the computational basis, showing large populations in the states $|00\rangle$ and $|11\rangle$. This state exhibits a nearly maximal negativity, with $\mathcal{N}(\rho_{78}) = 0.499$.}
    \label{fig: state_bell2_square}
\end{figure}

\begin{figure}[h]
    \centering
    \includegraphics[width=1\linewidth]
    {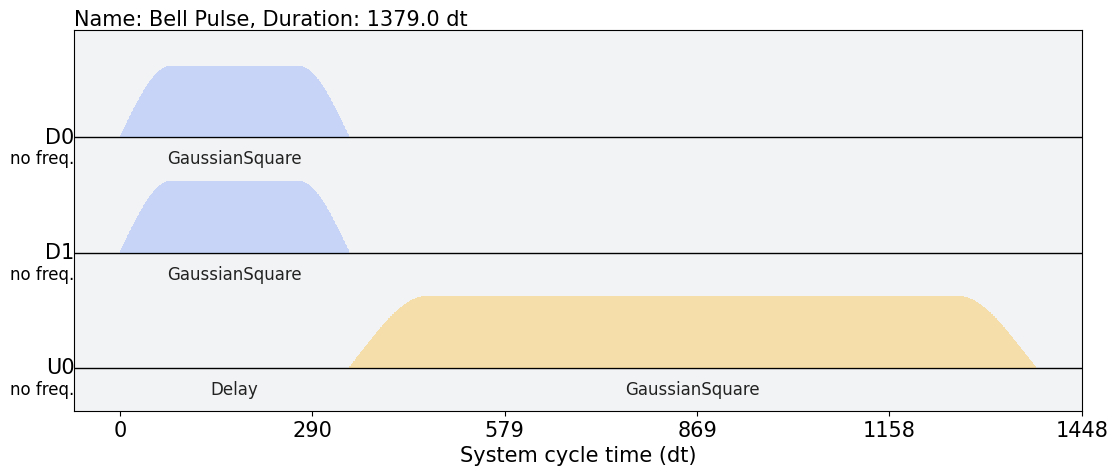}
    \caption{Gaussian Square pulse scheme used to obtain a Bell-type state between qubits 7 and 8, considering three energy levels. The pulse widths were fixed based on those obtained in the square-pulse case, shown in Fig.~\ref{fig: pulse_bell2_square}. In this way, only the amplitude and $\sigma$ parameters were varied, where $\sigma$ characterizes the smoothness of the envelope edges. The diagram also shows that the total execution time is $1379.0,dt$.}
    \label{fig: pulse_bell2_gaussian}
\end{figure}

In the implementation described above, we assumed that the system can be effectively described by a two-level model, under the approximation that the third energy level is sufficiently far detuned such that the probability of undesired excitation to higher levels is negligible. Nevertheless, in realistic physical systems, the presence of the third level cannot be ignored, since population leakage to this level may occur. For this reason, we consider a three-level model using the parameters described in Eqs.~\ref{eq_5} and \ref{eq_6}, as well as those presented in Section~\ref{imple}.
\\\\
In this case, we use the same pulse sequence as in the previous implementation, with the difference that the pulses are modeled using Gaussian Square functions, which allows us to capture the system dynamics more realistically (see Fig.~\ref{fig: pulse_bell2_gaussian}). The characteristic parameters of this function are the amplitude ($A$), width, standard deviation ($\sigma$), and duration, which are detailed in Appendix~\ref{Appendix C}. This sequence allows us to obtain a locally unitary (LU)-equivalent version of the Bell state, with a negativity of $\mathcal{N}(\rho_{01}) = 0.499$. For this implementation, the resulting state is given by Eq.~\ref{psi_gauss}.

%--------------------- fase global

It is worth noting that the widths of the Gaussian Square functions were fixed using the square pulses obtained in Fig.~\ref{fig: pulse_bell2_square}. In this way, it was only necessary to optimize the $\sigma$ parameter of each function. As a result of the evolution, the final state vector is described by nine components. Subsequently, the components associated with the third level were removed, the resulting state was renormalized, and the cost function was minimized to determine the optimal parameters.

\begin{figure}[H]
    \centering
    \includegraphics[width=1\linewidth]
    {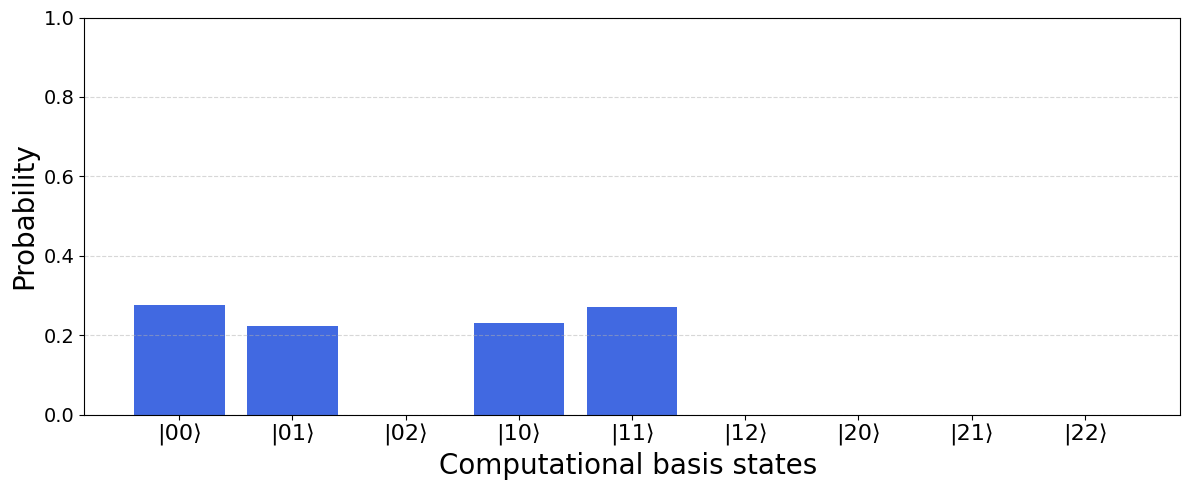}
    \caption{Probability distribution in the computational basis for the two-qubit state in Eq.~\ref{psi_gauss}. This distribution is obtained by applying the pulse sequence shown in Fig.~\ref{fig: pulse_bell2_gaussian} while retaining the first three energy levels of each qubit. Thus, the system is described by two qutrits. The negativity reaches $\mathcal{N}(\rho_{01}) = 0.499$ when the third levels are omitted.}
    \label{fig: state_bell2_gaussian}
\end{figure}

Finally, the population probabilities of the state vector were computed using the optimal parameters obtained. The results are presented in Fig.~\ref{fig: state_bell2_gaussian}, showing that the resulting populations are very similar.

\subsection{Tripartite entanglement}

In three-qubit systems, entanglement is divided into inequivalent SLOCC classes \cite{dur2000three}. Here, we use optimal pulse control to prepare states belonging to each class. For a given target class, we optimize the pulse parameters such that the final state is locally unitary (LU) equivalent to the corresponding canonical representative. In what follows, we outline the optimization procedure and report the results obtained for each entanglement class.

\vspace{0.5 cm}

Beyond the SLOCC approach, three-qubit states can also be classified using invariant quantities. These invariants are algebraic functions of the computational-basis coefficients and remain unchanged under local unitary (LU) transformations. They provide a direct way to identify the entanglement family without explicitly converting the state to a canonical form. A key example is the 3-tangle \cite{acin2001three}, it is nonzero only for the GHZ-class and vanishes for the W-class. In our analysis, we use such invariants to validate the output of the pulse optimization by comparing their values with those of the canonical representative of each target class. 
\subsubsection{GHZ state}

\begin{figure}[H]
    \centering
    \includegraphics[width=1\linewidth]
    {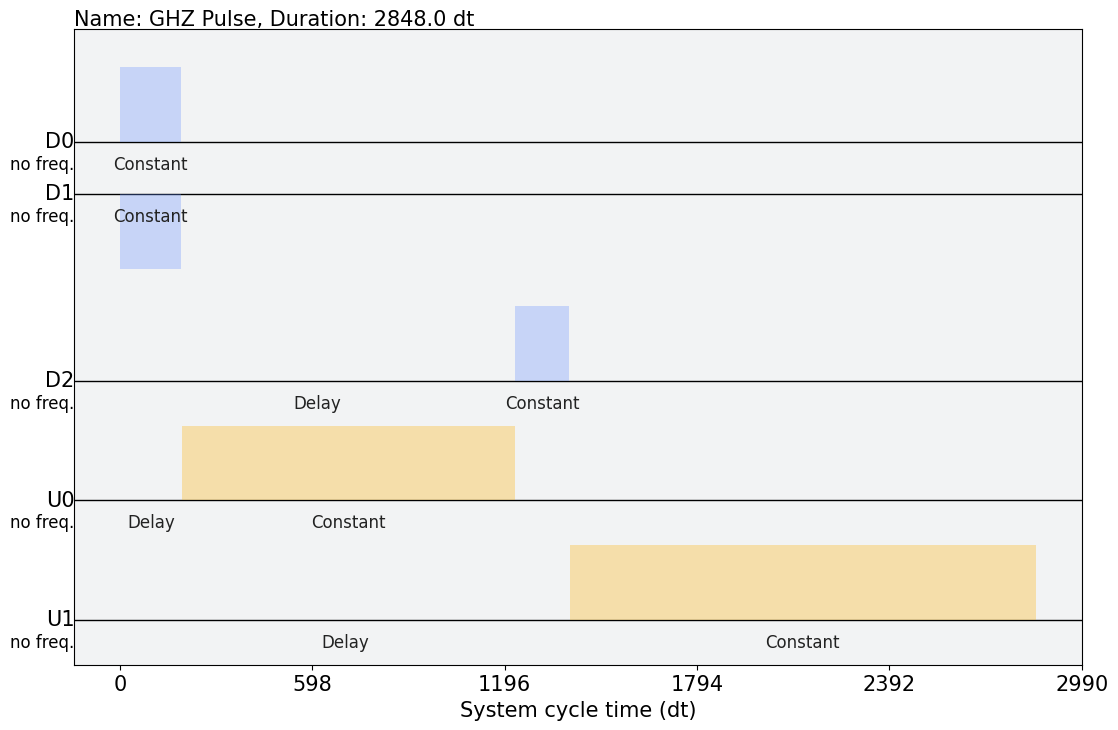}
    \caption{Square-pulse scheme used to generate a three-qubit state in the GHZ entanglement class. Three local drive pulses (cyan) are applied through channels D0, D1, and D2, while two cross-resonance control pulses (yellow) are applied through channels U0 and U1. Pulse U0 drives qubit 7 at the resonance frequency of qubit 8, whereas pulse U1 drives qubit 8 at the resonance frequency of qubit 9. These cross-resonance drives induce the effective interactions required to extend the correlations across the three-qubit system. The total duration of the pulse sequence is 2848.0~dt.}
    \label{fig: pulse_ghz_square}
\end{figure}

To prepare a Greenberger--Horne--Zeilinger (GHZ) class state in our system, we design a coherent microwave pulse protocol. The pulse scheme, shown in Fig. \ref{fig: pulse_ghz_square}, consists of five constant-amplitude pulses: three local drive pulses applied through the drive channels (D0, D1, and D2) and two cross-resonance control pulses (U0 and U1). Each pulse is characterized by its amplitude and duration, which are optimized to maximize the degree of tripartite entanglement.

\vspace{0.2 cm}

The sequence begins with two simultaneous drive pulses applied to qubits 7 and 8 through channels D0 and D1, respectively, with identical durations. These pulses initiate the coherent evolution of the two-qubit system. Subsequently, a control pulse is applied through channel U0, corresponding to a drive on qubit 7 at the resonance frequency of qubit 8. In the presence of the inter-qubit coupling, this cross-resonance drive induces an effective interaction between the two qubits, thereby generating quantum correlations between them.

\vspace{0.1 cm}

In the second stage, the third qubit is activated by a drive pulse applied to qubit 9 through channel D2. The sequence concludes with a control pulse on channel U1, which drives qubit 8 at the resonance frequency of qubit 9. In the presence of the inter-qubit coupling, this cross-resonance drive induces an effective interaction between qubits 8 and 9, extending the correlations established between qubits 7 and 8 to the three-qubit system.

The control parameters are optimized through an optimal-control search over the parameter space, using the \textit{three-tangle} $(\tau_{3})$ as the objective function. This quantity measures genuine three-qubit entanglement and reaches its maximum value for the GHZ state. In our implementation, the optimized value of the three-tangle reaches $\tau_{3} \approx 0.999$, making it an effective measure for guiding the optimization toward the GHZ entanglement class.

\begin{figure}[H]
    \centering
    \includegraphics[width=1\linewidth]
    {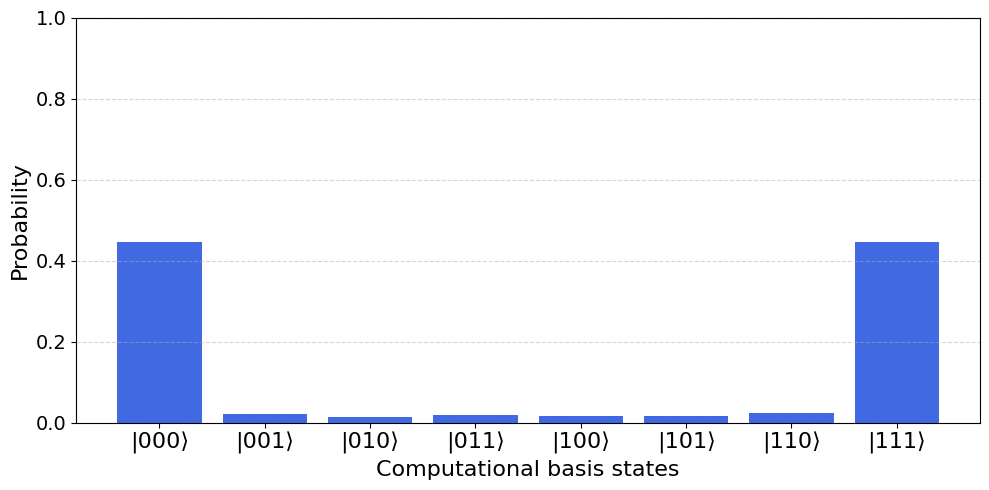}
    \caption{Probability distribution obtained from the pulse scheme shown in Fig.~\ref{fig: pulse_ghz_square}. The resulting state achieves a three-tangle of $\tau_{3} \approx 0.999$. The dominant populations correspond to $|000\rangle$ and $|111\rangle$, while the remaining computational-basis states have significantly smaller populations.}
    \label{fig: state_ghz_square}
\end{figure}

Using the pulse configuration introduced above, we obtain the probability distribution shown in Fig.~\ref{fig: state_ghz_square}. The dominant populations here correspond to $|000\rangle$ and $|111\rangle$, while the remaining basis states contribute weakly. This indicates a predominantly coherent superposition of $|000\rangle$ and $|111\rangle$. Small deviations from the ideal distribution reflect practical limitations of calibration and optimization, while preserving the genuinely tripartite nature of the generated entanglement.

With the pulse sequence represented in Figure \ref{fig: pulse_ghz_square}, we obtain the following GHZ state:

%--------Fase global:

\begin{equation}
\begin{split}
|\Psi'_{GHZ}\rangle
=\, &(0.66802070)\,|000\rangle\\
&+\,(0.05252552 + 0.13434712\, i)\,|001\rangle\\
&+\,(-0.10755963 + 0.03577012\, i)\,|010\rangle\\
&+\,(-0.11451824 + 0.07573796\, i)\,|011\rangle\\
&+\,(-0.03285393 - 0.11985692\, i)\,|100\rangle\\
&+\,(-0.06979810 - 0.10802873\, i)\,|101\rangle\\
&+\,(0.11835156 - 0.10244015\, i)\,|110\rangle\\
&+\,(-0.65590019 - 0.12061437\, i)\,|111\rangle.
\end{split}
\label{GHZ-cuadrado}
\end{equation}

\begin{figure}[H]
    \centering
    \includegraphics[width=1\linewidth]
    {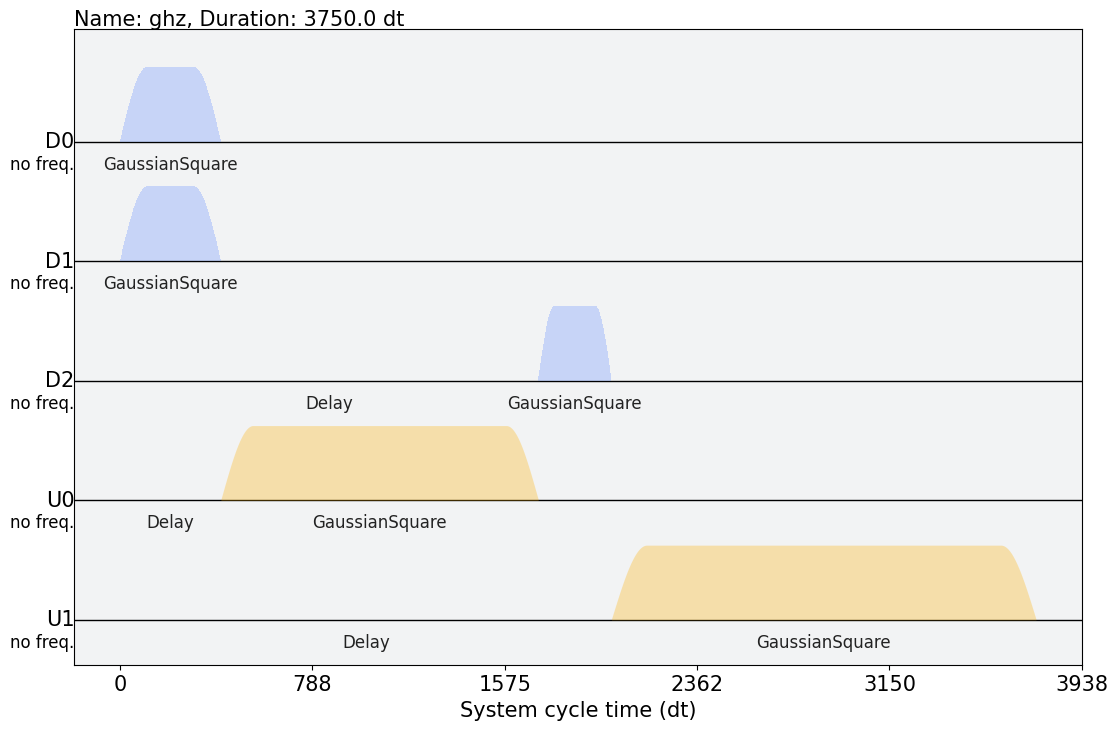}
    \caption{Pulse scheme used to generate a three-qubit state in the GHZ entanglement class when considering three energy levels for each qubit. The sequence includes three local drive pulses (cyan) applied through channels D0, D1, and D2, and two cross-resonance control pulses (yellow) applied through channels U0 and U1. The pulse parameters are optimized by maximizing the three-tangle. The widths of the Gaussian Square pulses are determined from the corresponding square-pulse scheme; therefore, the amplitude $A$ and standard deviation $\sigma$ are the optimization parameters for each pulse. The total duration of the pulse sequence is 3750.0~dt.}
    \label{fig: pulse_ghz_gaussian}
\end{figure}

As in the two-system case, we consider a more realistic model with three energy levels for each individual system. In this case, we solve the time-dependent Schr\"odinger equation using the pulse scheme shown in Fig.~\ref{fig: pulse_ghz_gaussian}. Each pulse is modeled using a Gaussian Square envelope, with the square-pulse scheme shown in Fig.~\ref{fig: pulse_ghz_square} serving as a reference. The widths of the square pulses are used to define the corresponding pulse durations in the Gaussian Square scheme. In this setting, we optimize both the pulse amplitudes and the standard deviation $\sigma$, again using the three-tangle as the objective function.

Since three systems with three levels per site evolve in a 27-dimensional state vector, at the end of the dynamics, we restore the three transmon states by discarding entries associated with the third level for each qubit. We then renormalize the state and evaluate the cost function. As a result, we obtain a three-tangle $\tau_{3}\approx0.999$.

\begin{figure}[H]
    \centering
    \includegraphics[width=1\linewidth]
    {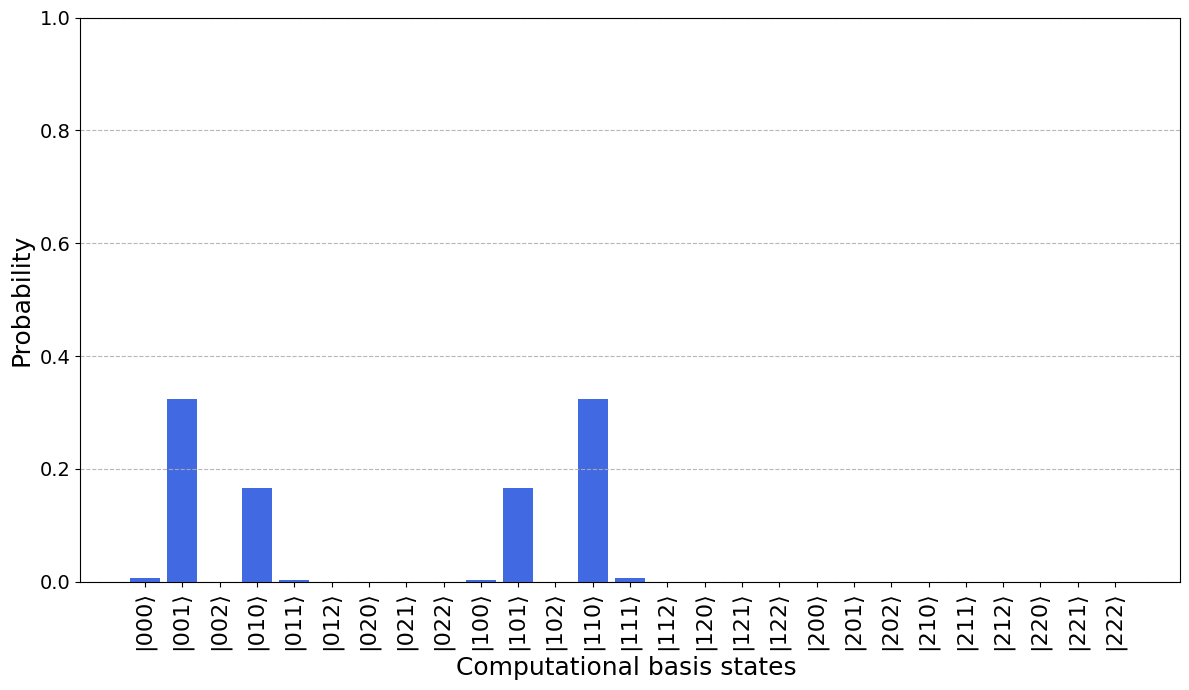}
    \caption{Probabilities of the computational basis state obtained by optimizing the pulse scheme in Fig.~\ref{fig: pulse_ghz_gaussian}. The results show that the states $|001\rangle$ and $|110\rangle$ have nearly populations, while $|010\rangle$ and $|101 \rangle$ also exhibit similar probabilities. The remaining basis states have negligible populations, particularly those involving the $|2\rangle$ level, indicating the population remains predominantly within the computational subspace.}
    \label{fig: state_ghz_gaussian}
\end{figure}

The probability distribution of the resulting state is shown in Fig.~\ref{fig: state_ghz_gaussian}. Four computational-basis states exhibit significant populations, with approximately balanced populations within certain pairs. Despite the presence of additional energy levels, the resulting state retains a high degree of tripartite entanglement, as quantified by the three-tangle. These results demonstrate the robustness of the proposed pulse protocol against the inclusion of higher energy levels.

The state obtained for the previous configuration is \ref{GHZ_gauss}, along with its corresponding probability distribution.

%----- Fase global

\subsubsection{W state}

\begin{figure}[H]
    \centering
    \includegraphics[width=1\linewidth]
    {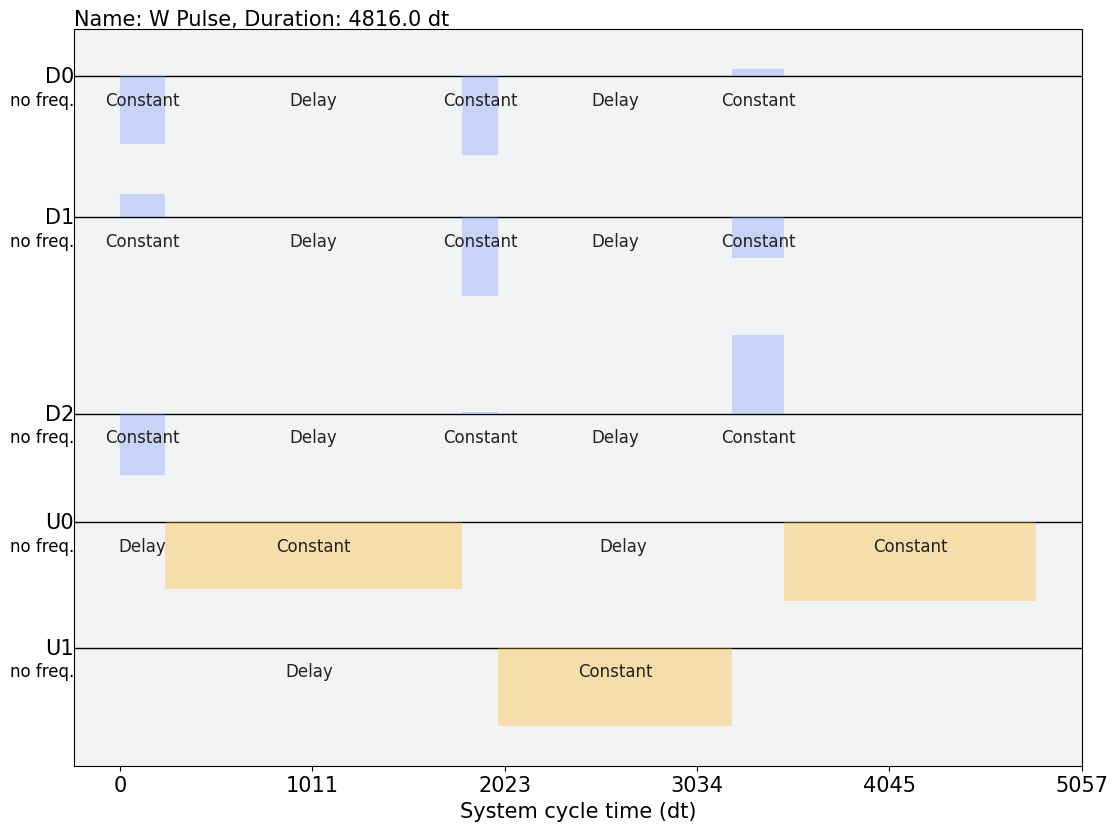}
    \caption{Pulse scheme used to obtain a $W$-class state in the two-level model. The sequence consists of nine local pulses (cyan) and three cross-resonance pulses (yellow). It begins with three equal-duration pulses applied to channels $D0$, $D1$, and $D2$, followed by a pulse on channel $U0$. A second set of three equal-duration pulses is then applied to $D0$, $D1$, and $D2$, followed by a cross-resonance pulse on channel $U1$. Subsequently, a third set of three equal-duration pulses is applied to the local drive channels, and the sequence concludes with a cross-resonance pulse on channel $U0$. The total duration of the pulse sequence is $4816.0,dt$.}
    \label{fig: pulse_w_square}
\end{figure}

In order to prepare a three-qubit $W$-class state, we design the coherent pulse protocol shown in Fig. \ref{fig: pulse_w_square}. The sequence consists of constant-amplitude pulses applied to the local drive channels and the cross-resonance control channels ($U0$ and $U1$). Each pulse is characterized by its amplitude and duration, which are jointly optimized to maximize the desired entanglement properties of the resulting state.

The pulse sequence begins with the application of pulses to the three qubits through the local drive channels $D0$, $D1$, and $D2$, establishing the initial coherent dynamics of the system. Subsequently, a control pulse is applied through channel $U0$, activating the interaction between qubits 7 and 8 through a cross-resonance drive and contributing to the generation of bipartite correlations within the subsystem.
\\
In the second stage, three additional excitation pulses are applied to the local drive channels $D0$, $D1$, and $D2$. Subsequently, a control pulse is applied to channel $U1$ to induce an interaction between qubits 8 and 9, thereby extending the correlations to the third qubit. Finally, three additional pulses are applied to the local drive channels $D0$, $D1$, and $D2$, followed by a pulse on channel $U0$ to further enhance the interaction between qubits 7 and 8.
\\\\
The optimization is based on the squared pairwise concurrences, with the theoretical value of $\frac{4}{9}$ for each pair serving as the target value. The resulting state, which provides the best approximation to a state in the $W$ entanglement class according to this cost function, is given by

%-------------- Fase global

\begin{equation}
\begin{split}
|\Psi'_W\rangle
=\, &(0.51397601)\,|000\rangle\\
&+\,(-0.10295038 + 0.20024143\, i)\,|001\rangle\\
&+\,(-0.17617122 - 0.12098743\, i)\,|010\rangle\\
&+\,(0.30108107 - 0.16458899\, i)\,|011\rangle\\
&+\,(-0.20955602 - 0.12563232\, i)\,|100\rangle\\
&+\,(-0.35878680 + 0.39267187\, i)\,|101\rangle\\
&+\,(0.17604487 + 0.13212849\, i)\,|110\rangle\\
&+\,(-0.30967882 + 0.18642040\, i)\,|111\rangle.
\end{split}
\label{W-cuadrado}
\end{equation}

\begin{figure}[H]
    \centering
    \includegraphics[width=1\linewidth]
    {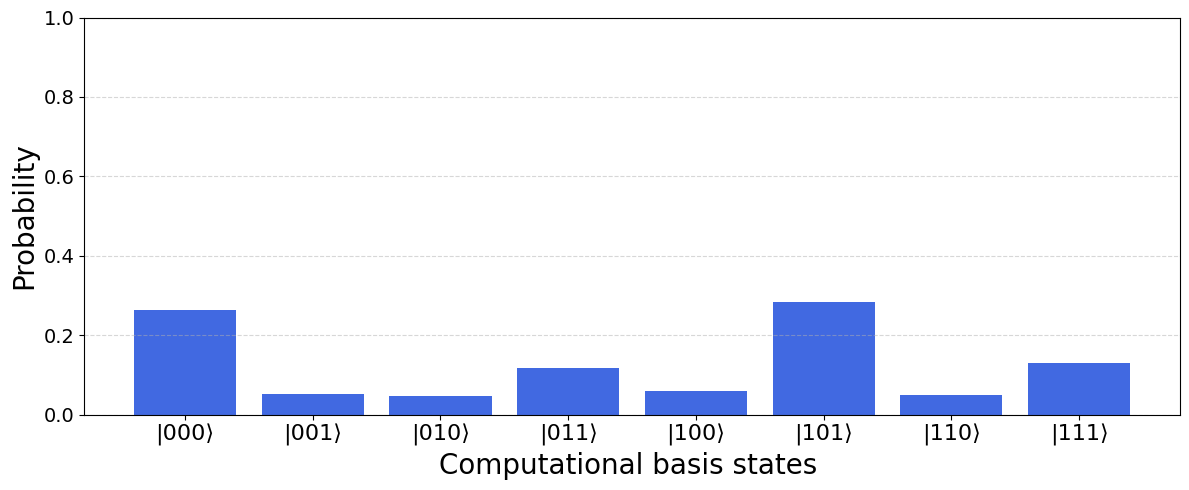}
    \caption{Probability distribution in the computational basis obtained by applying the optimized pulse sequence shown in Fig.~\ref{fig: pulse_w_square}. The population is distributed among the computational basis states, with the largest contributions corresponding to $|000\rangle$ and $|101\rangle$. The resulting squared pairwise concurrences are $\mathcal{C}^{2}(\rho_{01}) = 0.442$, $\mathcal{C}^{2}(\rho_{02}) = 0.444$, and $\mathcal{C}^{2}(\rho_{12}) = 0.442$, indicating a similar degree of bipartite entanglement among the three qubit pairs and a close agreement with the ideal value of $4/9$.}
    \label{fig: state_w_square}
\end{figure}

\begin{table}[H] % O [htbp]
\centering % Comando para centrar la tabla
\begin{tabular}{c|c} 
Sq. Conc. & Value \\ 
\hline 
$\mathcal{C}^{2}(\rho_{01})$ & 0.444 \\ 
$\mathcal{C}^{2}(\rho_{02})$ & 0.442 \\  
$\mathcal{C}^{2}(\rho_{12})$ & 0.442  
\end{tabular} 
\caption{Squared concurrences for the $W$ state optimization.} 
\label{tab: concurrence_square} 
\end{table}

These results in the table \ref{tab: concurrence_square} show that each bipartition exhibits nearly the same amount of quantum correlation. This is a hallmark of W-class entanglement, where losing one qubit does not destroy the correlations among the remaining qubits.

As a result of the optimized protocol, we obtain the probability distribution shown in Fig.~\ref{fig: state_w_square}. The highest populations are observed for the states $|000\rangle$ and $|101\rangle$, followed by the nearly equal populations of $|011\rangle$ and $|111\rangle$. The remaining basis states exhibit smaller populations, which can be attributed to residual population leakage and unwanted dynamical effects arising from the system's interactions. A comparison between the resulting squared pairwise concurrences and their ideal values is presented in Table~\ref{tab: concurrence_square}.

\begin{table}[H] % O [htbp]
\centering % Comando para centrar la tabla
\begin{tabular}{c|c|c|c} 
Squared concurrence & $W_{num}$ & $W_{ide}$ & Error $(\% )$    \\ 
\hline 
$\mathcal{C}^{2}(\rho_{01})$ & 0.4443 & 0.4444  & 0.032\\ 
$\mathcal{C}^{2}(\rho_{02})$ & 0.4422 & 0.4444 & 0.505\\ 
$\mathcal{C}^{2}(\rho_{12})$ & 0.4420 &  0.4444 & 0.550\\ 
\end{tabular} 
\caption{Comparison between the squared concurrences of the obtained state and the ideal W state.} 
\label{tab: error_w} 
\end{table}

Table~\ref{tab: error_w} compares the squared pairwise concurrences of the numerically generated state with those of the ideal $W$ state. The numerical results closely agree with the corresponding ideal values, with relatively small discrepancies for the three reduced two-qubit states. This agreement indicates that the numerically generated state reproduces the pairwise entanglement properties of the ideal $W$ state with high accuracy, suggesting that the two states may be related by local unitary transformations.
\\\\
To further investigate this equivalence, an independent analysis was performed using the Bures distance ($D_B$) to determine the local unitary transformations that optimally map the numerically generated state onto the ideal $W$ state. The parameters of the three local unitary transformations were obtained through an optimization procedure in which the Bures distance was used as the cost function, employing the Nelder--Mead algorithm. The resulting Bures distance was $D_B=0.056$, indicating a small separation between the two states after optimization and providing further evidence that the numerically obtained state is consistent with the $W$ entanglement class up to local unitary transformations (see Appendix~\ref{apen Bures Distance}).

\begin{figure}[h]
    \centering
    \includegraphics[width=1\linewidth]
    {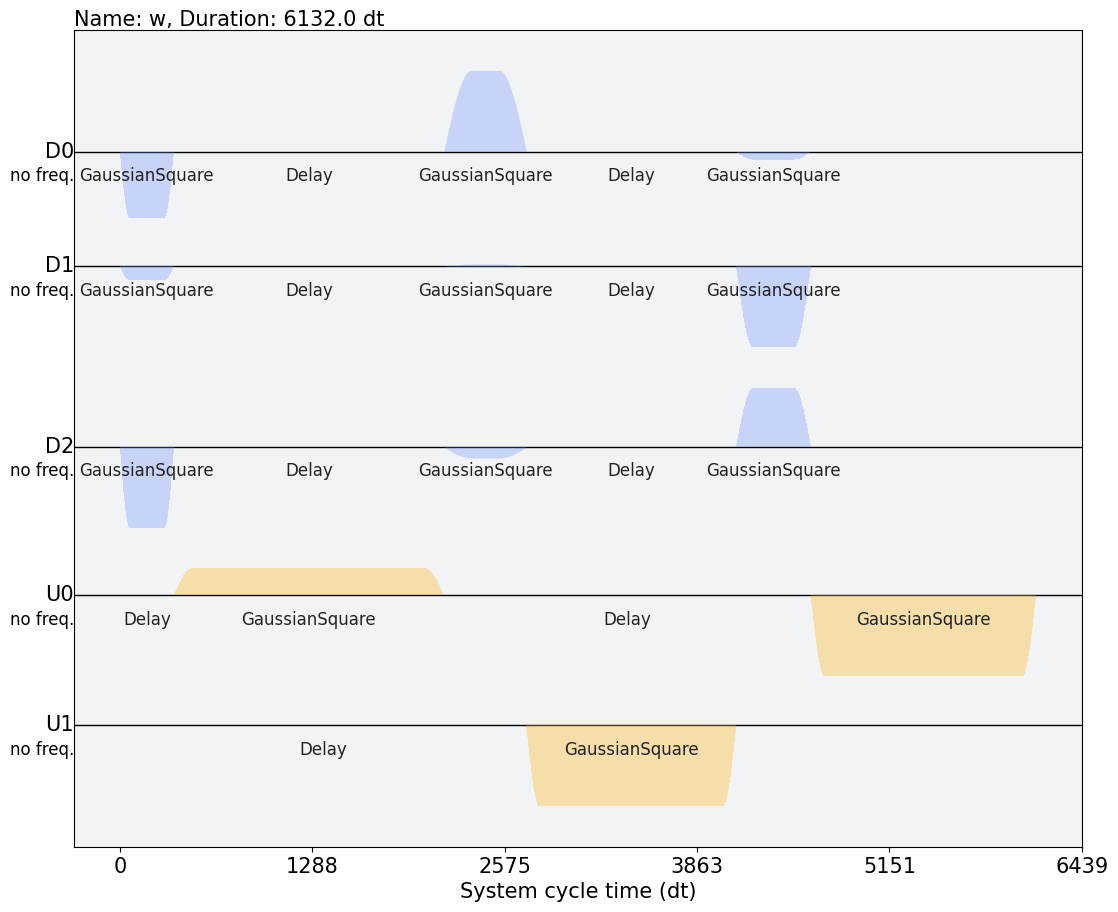}
\caption{Pulse scheme used to obtain a state equivalent to a $W$ state. The sequence includes 12 Gaussian Square pulses, consisting of nine local pulses (light blue), which are applied directly to the individual qubits, and three cross-resonance pulses. The width of each pulse is set based on the square-pulse results (see Figure). The optimized parameters are the amplitude of each pulse and $\sigma$, which controls the edge smoothness of each envelope. The total duration of the pulse scheme was 6132.0 $dt$.
}
    \label{fig: pulse_w_gaussian}
\end{figure}

To complete the study, we consider a three-qubit system with three levels per qubit. The dynamics are obtained by integrating the time-dependent Schr\"odinger equation using the pulse scheme shown in Fig.~\ref{fig: pulse_w_gaussian}. The implemented pulses are Gaussian Square envelopes arranged according to the same temporal pattern as the square pulses shown in Fig.~\ref{fig: pulse_w_square}. The total duration of each pulse is therefore fixed by the previously optimized duration of the corresponding square pulse, while the free parameters are the pulse amplitudes and the $\sigma$ values that determine the Gaussian envelopes.

In this implementation, the three levels of each system are explicitly included in the simulation. The final state is then projected onto the computational two-level subspace and renormalized before evaluating the cost function.
\\
With this configuration, we obtain the following squared concurrence shown in Table \ref{table3}:
\begin{table}[H] % O [htbp]
\centering % Comando para centrar la tabla
\begin{tabular}{c|c} 
Sq. Conc. & Value \\ 
\hline 
$\mathcal{C}^{2}(\rho_{01})$ & 0.441 \\ 
$\mathcal{C}^{2}(\rho_{02})$ & 0.435 \\  
$\mathcal{C}^{2}(\rho_{12})$ & 0.435  
\end{tabular} 
\caption{Squared concurrence values obtained in the optimization of the W state for the Gaussian Square implementation.} 
\label{table3} 
\end{table}

\begin{figure}[h]
    \centering
    \includegraphics[width=1\linewidth]
    {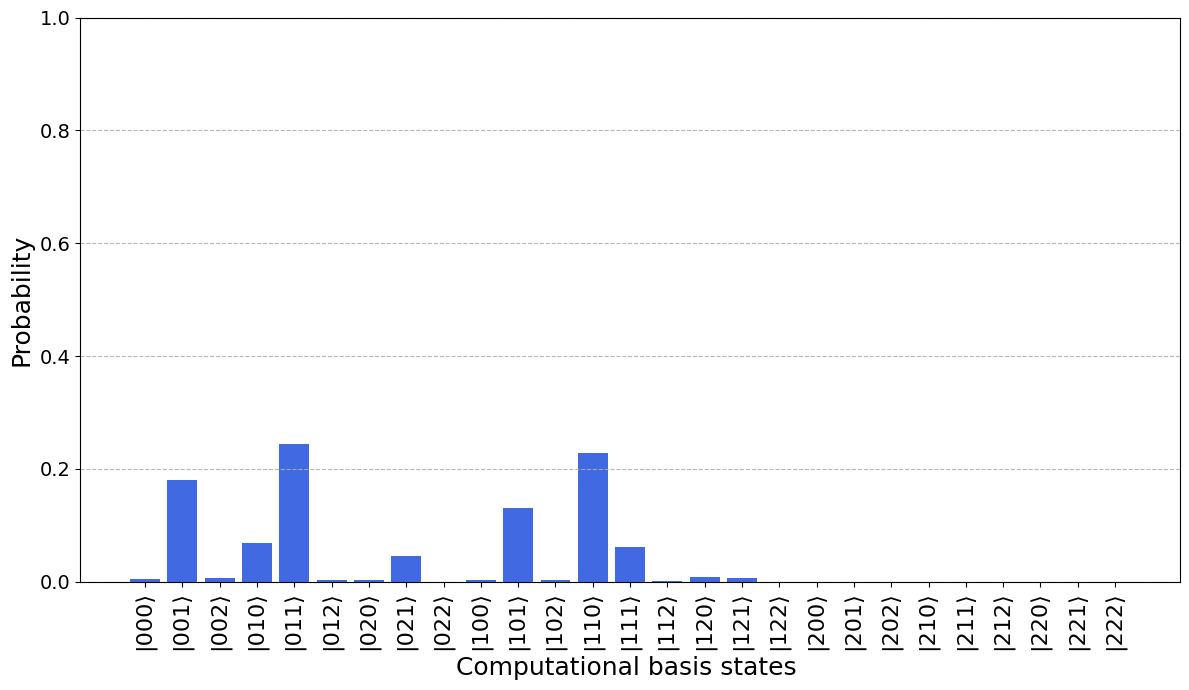}
\caption{Probability distribution over the computational basis states for the state (\ref{eq: gauss_w}) prepared with the Gaussian pulse scheme shown in Fig.~(\ref{fig: pulse_w_gaussian}). The histogram shows the measured probability associated with each basis state following the pulse sequence. While most of the population remains within the intended computational subspace, leakage to higher-energy states is also observed.}
    \label{fig: state_W_gaussian}
\end{figure}

These values reveal a close approach to the theoretical W-state benchmark. Figure \ref{fig: state_W_gaussian} shows that the final probability distribution is concentrated on seven of the twenty-seven computational basis states, with larger weights on $|011\rangle$ and $|110\rangle$. This provides clear evidence that the protocol generates a W-like quantum state. Leakage to the third energy level can also be observed in some states.

%\begin{equation}
%    \begin{split}
%        |\Psi_{w\_gauss}\rangle = (0.0721809805 + 0.0153940039 i) |000\rangle\\
%        +(0.4413136053 - 0.0251241402 i) |001\rangle\\
%        +(0.2571692968 - 0.094443977 i) |010\rangle\\
%        -(0.4101257074 - 0.3120861425 i) |011\rangle\\
%        -(0.019074928 + 0.0525841299 i) |100\rangle\\
%        -(0.1822897495 + 0.3282705161 i) |101\rangle\\
%        -(0.3277610058 + 0.3733210332 i) |110\rangle\\
%        -(0.2469472538 + 0.0812932767 i) |111\rangle
%    \end{split}
%\end{equation}

%------------ Fase global: 

\begin{figure}[H]
    \centering
    \includegraphics[width=1\linewidth]
    {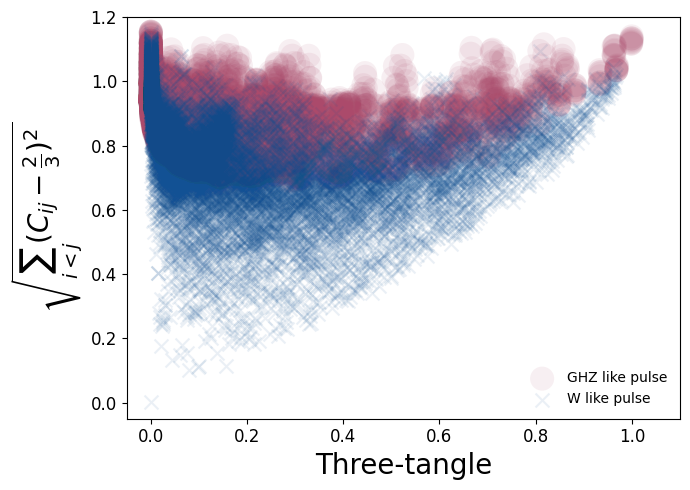}
\caption{Distribution of bipartite and tripartite entanglement measures for states generated by the GHZ- and W-like pulse schemes. The quantity $(\sqrt{\sum_{i<j}(C_{ij}-2/3)^2})$, where $C_{ij}$ denotes the concurrence of the bipartite reduced state $\rho_{ij}$, is plotted as a function of the three-tangle $\tau$. The purple circles and blue crosses correspond to states generated using the pulse schemes shown in Figs.~\ref{fig: pulse_ghz_square} and \ref{fig: pulse_w_square}, respectively, while varying the pulse amplitudes and durations. The parameter sweep explores the range of bipartite and tripartite entanglement measures accessible with each pulse protocol. The overlap between the two distributions indicates regions in which the GHZ- and W-like pulse schemes generate states with similar combinations of bipartite and tripartite entanglement.}
    \label{fig: correlation measures} 
\end{figure}

The figure \ref{fig: correlation measures} reveals a substantial overlap between the correlations generated by the GHZ and W-type pulse schemes, demonstrating that appropriate variations in pulse amplitude and duration can lead to states with comparable combinations of bipartite concurrence and tripartite entanglement. However, the two distributions are not identical and do not correspond one-to-one, indicating that the pulse protocols maintain distinct entanglement structures. In particular, the overlap demonstrates that similar values of tripartite entanglement can be accompanied by different distributions of bipartite correlations.

\begin{center}
\begin{table}[H]
\begin{tabular}{c|c|c|c|c|c|c}
State & Gates scheme & $t_{gates}$ & square pulse scheme & $t_{pulses}$ \\
\hline
Bell &  Figure \ref{fig: transpile_IBM_sherbrook_circuit_Bell} & 2912 $dt$ & Figure \ref{fig: pulse_bell2_gaussian} & 1379.0 $dt$\\
GHZ  &  Figure \ref{fig:transpile_IBM_sherbrook_circuit_GHZ}   & 5315 $dt$ & Figure \ref{fig: pulse_ghz_gaussian} & 3750.0 $dt$\\
W    &  Figure \ref{fig:transpile_IBM_sherbrook_circuit_w}     & 8224 $dt$ & Figure \ref{fig: pulse_w_gaussian} & 6132.0 $dt$\\
\end{tabular}
\caption{The execution time of the gate sequences required to implement representative quantum states, such as Bell, GHZ, and W states, is compared with the time required by our pulse schemes to achieve the same quantum correlations. In this sense, the states generated by our pulse schemes differ from the corresponding reference states by, at most, local unitary operations. The execution times of both the gate-based and pulse-based implementations (see Appendix~\ref{APPEN A}) are taken from the IBM Sherbrooke quantum processor, for which the discrete time step is (dt = 0.2222) ns. In the pulse-based implementation, the control pulses are modeled using Gaussian-square functions.}
\label{tabla4}
\end{table}
\end{center}

\section{Simulations with decoherence}

\begin{figure}[H]
    \centering
\includegraphics[width=0.9\linewidth]{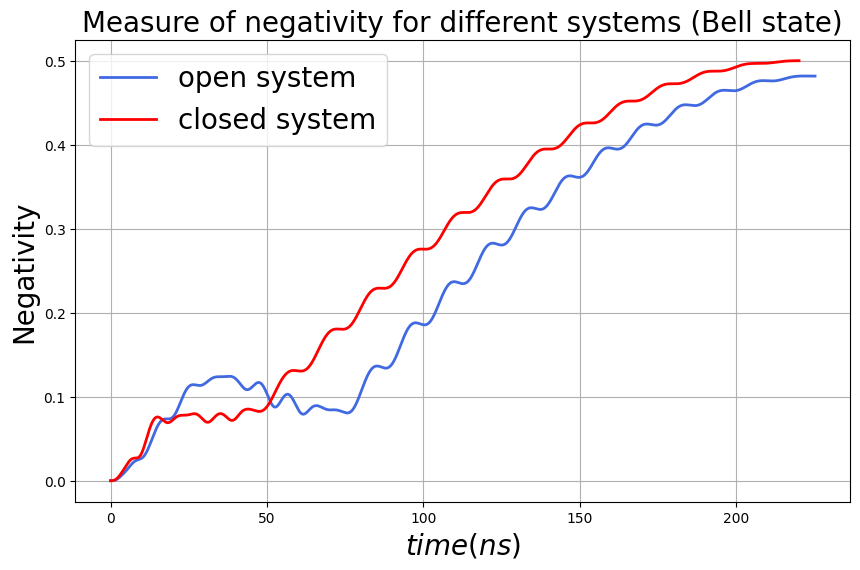}
    \caption{ Negativity as a function of time for the pulse scheme presented in  Fig. \ref{fig: pulse_bell2_square}. considering both closed (red) and open (blue) system dynamics. For the closed system, the optimized dynamics exhibits a smooth overall increase in negativity, reaching a maximum value of (0.4999), which is close to the maximum value of (0.5) for a maximally entangled Bell state. Small oscillations can be observed throughout the evolution, reflecting the coherent dynamics of the system. In the open system, the negativity also increases with time but exhibits more pronounced oscillations and reaches a slightly lower maximum value of (0.4831), as a result of the interaction with the environment and the associated decoherence effects.}
    \label{fig: negativi_lindblad}
\end{figure}

During the first approximately $(50,\mathrm{ns})$, the dynamics is dominated by the local drive channels D0 and D1. Although these are local control fields, they do not result in a strictly local unitary evolution because the static exchange interaction $J_{78}(b_0 b_1^\dagger+b_0^\dagger b_1)$ in Eq.~\ref{eq_5} remains active throughout the sequence. Starting from $|00\rangle$,the exchange interaction alone does not generate entanglement. However, the local drives create coherent superpositions with nonzero amplitudes in $| 01\rangle$ and $|10\rangle$, allowing the always-on exchange interaction to couple these states. Consequently, a finite amount of bipartite entanglement develops before the cross-resonance pulse is applied. From approximately $50\,\mathrm{ns}$ onward, the cross-resonance drive further enhances the entangling dynamics, leading to the main increase in negativity observed in Fig.~\ref{fig: negativi_lindblad}.

From approximately $(50,\mathrm{ns})$ onward, a resonant pulse is applied, allowing for controlled interaction between qubits 7 and 8 and resulting in a more pronounced increase in negativity in both cases. However, the response of the open system is weaker than that of the closed system, demonstrating the effect of the interaction with the environment on the generation and, particularly, the preservation of quantum correlations.

Furthermore, the oscillations observed in the response of the closed system were analyzed by fitting the time-dependent dynamics. The parameters obtained from these fits are presented in \ref{Numerical data for the open system in two qubits}. These oscillations can be attributed to the coherent dynamics of the system, during which the coupling between the qubits leads to periodic excitation transfer and, consequently, to a temporal modulation of the quantum correlations.

The early-time entanglement should not be interpreted as arising from a nearly resonant bare exchange interaction between qubits 7 and 8. Using ordinary frequencies, the qubit detuning is $\delta=|\omega_7-\omega_8|/(2\pi)=56.9437\,\mathrm{MHz}$, whereas the coupling strength is $J_{78}/(2\pi)=2.013\,\mathrm{MHz}$, giving $\delta/J_{78}\approx 28.3$. The bare exchange interaction is therefore appreciably off-resonant. During the initial stage, however, both qubits are driven simultaneously. Combining the drive scales reported in \ref{tab: ajuste -frecue} with the optimized amplitudes of the initial pulses yields effective drive scales of approximately $43\,\mathrm{MHz}$ and $86\, \mathrm{MHz}$, which are comparable to the qubit detuning. Thus, the early-time entanglement should be understood as a driven-coupled effect: the local drives create and reshape the coherent amplitudes, while the always-on exchange interaction provides the nonlocal component of the dynamics.
%%%%%%%%%%%%%%%%%%%%%%%%%%%%%%%%%%%%%%%%%%%%%%%%%%%%%
\begin{figure}[H]
    \centering
    \includegraphics[width= 1\linewidth]{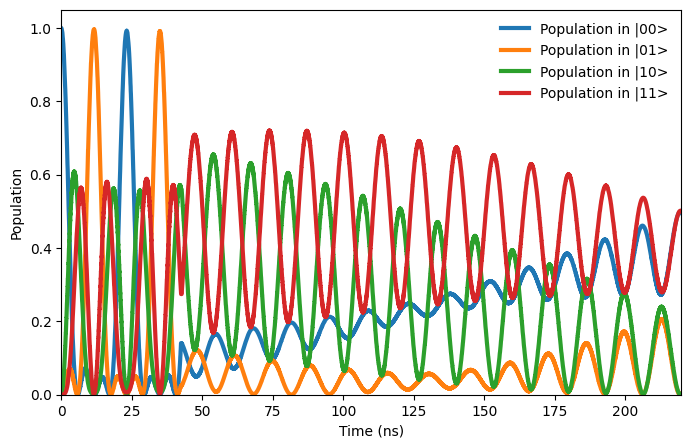}
    \caption{The oscillatory population dynamics reflect the coherent exchange of excitations between the qubits, mediated by the coupling $J_{78}$ and influenced by the external driving fields. The redistribution of populations among the computational basis states accompanies the generation and evolution of quantum correlations, thereby contributing to the oscillatory behavior observed in the negativity.}
    \label{fig: population_2q}
\end{figure}

The oscillatory behavior observed in the negativity is closely related to the coherent oscillations of the populations of the computational basis states. As shown inn Fig. \ref{fig: population_2q}, The populations of $|00\rangle$ , $(|01\rangle)$, $(|10\rangle)$, and $(|11\rangle)$ undergo periodic temporal variations due to the coherent exchange of excitations between the coupled qubits. These population oscillations are accompanied by corresponding variations in the off-diagonal coherences of the density matrix, which are essential for the evolution of quantum correlations. Consequently, the oscillatory component observed in the negativity can be associated with the underlying coherent population-transfer dynamics, while its overall increase results from the combined action of the qubit-qubit coupling and the external driving fields.

The net growth of negativity results from the combined driven and coupled evolution. The local drives create and reshape the coherent amplitudes on which the always-on interaction acts, while the interqubit  coupling $J_{78}$ provides the nonlocal component required for entanglement generation. The drives therefore modulate and enable the entangling action of the coupling rather than generating entanglement independently.

%%%%%%%%%%%%%%%%%%%%%%%%%%%%%%%%%%%%%%%%%%%%%%%%%

At this point, we consider the simulation of equation \ref{lindblad} for three coupled qubits, where the system Hamiltonian is given by the sum of Eqs. \ref{eq_7} and \ref{eq_8}. The parameters of the control pulses were taken from those obtained in the closed-system simulation ( see fig. \ref{fig: pulse_ghz_square}). The density matrix corresponding to the initial state $|000\rangle\langle000|$ was used as the initial condition, together with the decoherence parameters described in  \ref{Master equation implementation}.

The Genuine Multipartite Negativity (GMN) was evaluated for the quantum states obtained throughout the time-evolution simulation. This measure was used to quantitatively characterize the genuine multipartite entanglement generated in the system and to analyze its evolution under the applied control sequence. Specifically, the GMN was calculated at each time step from the corresponding density matrix, allowing the dynamics of multipartite entanglement to be characterized throughout the evolution. The resulting GMN values were then plotted as a function of time to identify the regime of entanglement generation and determine the maximum degree of genuine multipartite entanglement achieved during the evolution.

\begin{figure}[H]
    \centering
    \includegraphics[width=0.9\linewidth]{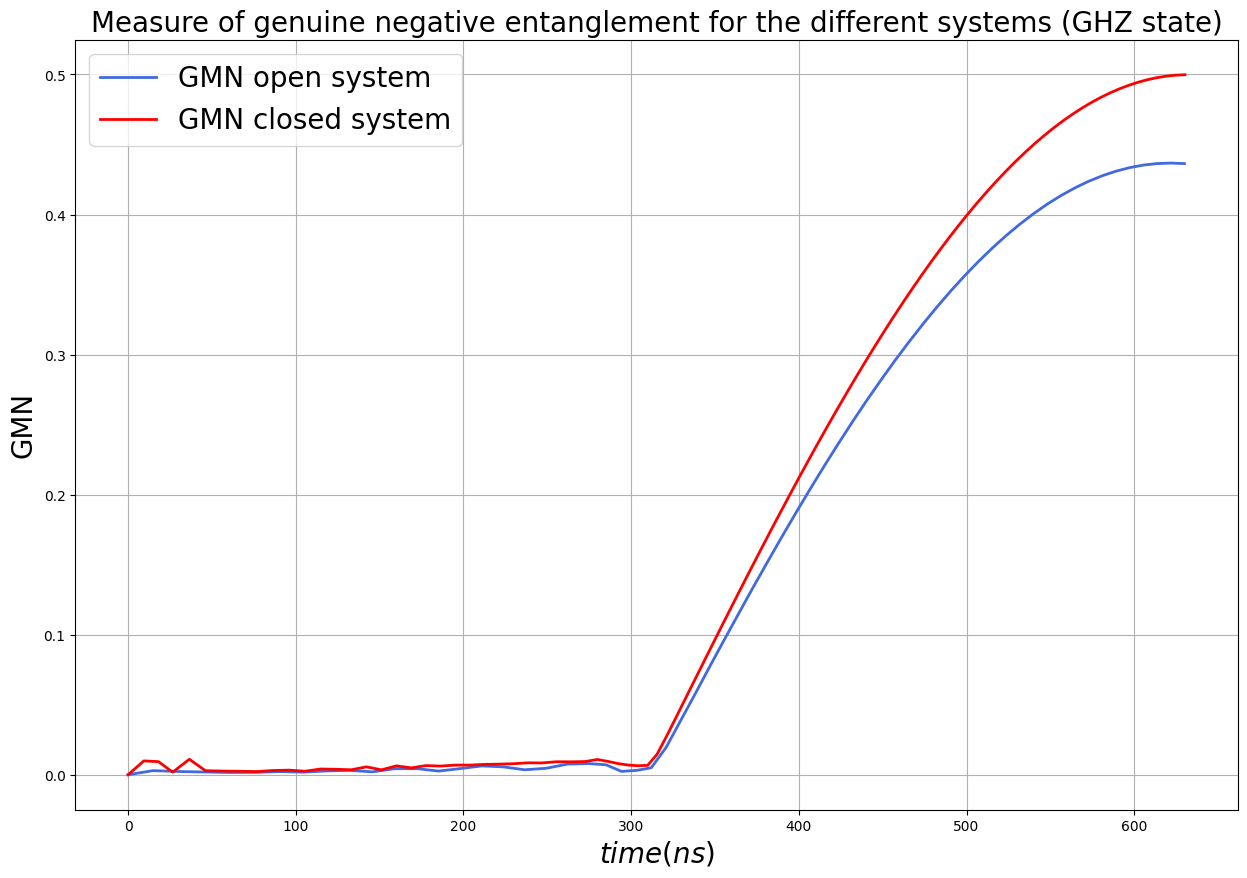}
    \caption{Comparison of the Genuine Multipartite Negativity (GMN) as a function of time for the GHZ pulse scheme. The blue curve represents the open-system evolution, in which decoherence effects are incorporated through the Lindblad master equation, while the orange curve represents the ideal closed-system evolution. Both systems exhibit similar dynamics during the initial stages; however, decoherence progressively reduces the generated multipartite entanglement, resulting in a maximum GMN of approximately $0.433$ for the open system, compared with $0.499$ for the closed system.}
    \label{fig: 3_tangle_tiempo}
\end{figure}

\section{Discussion}
\label{sec:discussion}

The results presented in this work demonstrate that optimized electromagnetic pulse sequences provide a viable alternative to gate-based strategies for controlling and generating entanglement in superconducting architectures. By directly targeting entanglement measures rather than specific target states, the proposed approach reduces the control overhead associated with state-specific preparation and enables the efficient generation of states exhibiting the characteristic entanglement properties of different multipartite classes.

This reduction in the effective expressivity of the control scheme can play an important role in mitigating error accumulation and decoherence effects. By restricting the accessible dynamical pathways to those required to generate a given quantum resource, the accumulation of control imperfections associated with long sequences of elementary gates can be reduced. Beyond noise mitigation, a reduced control expressivity may also offer practical advantages for optimization. In variational quantum algorithms, circuit expressivity has been linked to the emergence of barren plateaus, characterized by gradients that vanish exponentially with increasing system size \cite{holmes2022connecting}. Therefore, restricting the accessible control space may help avoid unnecessarily complex optimization landscapes while retaining the dynamical resources required to generate the desired entanglement properties. In this context, it is worth emphasizing that variational approaches formulated directly at the pulse level, in which logical gates are replaced by optimized electromagnetic controls, have already been successfully implemented on superconducting platforms.\cite{de2023pulse,wang2024pulse,liang2023hybrid,hansen2023pulse,egger2023pulse,sherbert2025parametrization}.

A notable feature of the proposed methodology is that the resulting pulse schemes are substantially shorter than those obtained by compiling standard quantum circuits into native gates. Even in the absence of explicit decoherence modeling, this reduction in total evolution time is relevant because shorter control sequences reduce the time during which the system is exposed to decoherence. These findings suggest that pulse-level control may provide a practical approach for mitigating the impact of decoherence in near-term superconducting processors while directly targeting the entanglement properties of interest.

From a broader perspective, the results support the idea that quantum optimal control can be naturally formulated at the level of quantum resources rather than at the level of unitary gates. By focusing on entanglement measures as optimization targets, the approach bypasses unnecessary constraints associated with circuit decompositions and opens the possibility of exploring otherwise inaccessible regions of the state space within realistic hardware limitations.

\section{Conclusions}
\label{sec:conclusions}

In this work, we have investigated quantum state preparation in superconducting transmon systems using optimized electromagnetic pulse sequences, moving beyond conventional gate-based paradigms. By directly optimizing entanglement measures, we have shown that both bipartite and multipartite quantum correlations can be efficiently generated with short, physically feasible control protocols.

For two-qubit systems, pulse sequences maximizing negativity yield states locally equivalent to maximally entangled Bell states, with their entanglement properties remaining robust when higher transmon levels are included. For three-qubit systems, we demonstrated the preparation of states belonging to the GHZ and W entanglement classes, using cost functions tailored to genuine tripartite correlations and balanced bipartite entanglement, respectively.

Beyond state preparation, the results presented in this work are directly relevant to the implementation of variational quantum algorithms on superconducting platforms. By operating at the pulse level and explicitly reducing the effective expressivity of the control scheme, the proposed approach naturally constrains the accessible region of the parameter landscape. Such restrictions are known to improve trainability by mitigating barren plateau phenomena, which arise when highly expressive parametrized circuits exhibit exponentially suppressed gradients. In this sense, pulse-based variational formulations provide a physically motivated mechanism to balance expressivity and optimization stability, suggesting that resource-oriented pulse control may play a key role in scalable variational algorithms on near-term quantum hardware.

Looking forward, the framework presented here can be extended to larger systems, different hardware architectures, and more complex resource measures. Incorporating decoherence models, robustness constraints with respect to Hamiltonian parameter uncertainties and control-amplitude fluctuations, and experimental calibration feedback constitutes a natural next step toward the experimental realization of resource-driven quantum state preparation protocols.
\\\\

\textbf{Acknowledgements}
We gratefully acknowledge the valuable comments provided by Herbert Díaz Moraga and Cristóbal Lledó Veloso. This work was supported by FONDECyT Regular nr 1230586, Chile. K. D. D. belongs to the Ph.D. program  \textit{Doctorado en física, mención Física-Matemática}, Universidad de Antofagasta, Chile, and acknowledges financial support from the ANID doctoral fellowship No. 21252201. Artificial intelligence tools (Claude and Fable 5.1) were used to improve the writing style and optimize the Python code.

\appendix
\section{Pulse-level implementation of gate-based circuits}
\label{APPEN A}

Quantum algorithms are implemented using quantum circuits composed of quantum gates. In superconducting quantum processors, these gates are realized through the precise control of electromagnetic pulses applied to the circuits. In platforms such as those developed by IBM, the hardware typically relies on a fixed set of quantum gates, each associated with predefined pulse shapes. Within these control schemes, cross-resonance pulses\cite{georgefrancis2025data,nguyen2024reinforcement,fischer2023universal} are employed to mediate interactions between pairs of superconducting circuits. Despite their effectiveness, such gates are susceptible to various sources of error, including crosstalk, leakage to non-computational states, and decoherence effects \cite{nguyen2024reinforcement,gautier2025optimal}. To mitigate these limitations, alternative pulse schemes, such as echo cross-resonance (ECR) \cite{sundaresan2020reducing}, have been extensively studied.
\\
The following figures show the quantum gates schemes that represented the basic structure to build the pulses schemes defined in this work. Here, we considered native gates from the IBM$\_$sherbrooke quantum hardware.

\begin{figure}[H]
    \centering
    \includegraphics[width=0.7\linewidth]
    {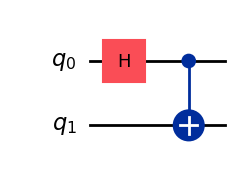}
\caption{Circuit representation of the Bell state $|\Psi\rangle = \frac{1}{\sqrt{2}}\left(|00\rangle + |11\rangle \right)$,obtained using a Hadamard gate followed by a CNOT gate, where $q_{0}$ is the control qubit and $q_{1}$ is the target qubit. }
    \label{fig: circuit_Bell_state}
\end{figure}

\begin{figure}[H]
    \centering
    \includegraphics[width=1\linewidth]
    {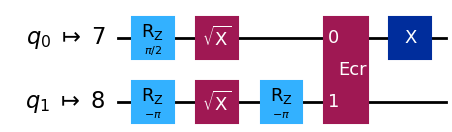}
\caption{Representation of the transpiled circuit shown in\ref{fig: circuit_Bell_state} using the native gates of the IBM$\_$sherbrooke simulator. Qubit seven is the control qubit, while qubit eight is the target qubit. In this case, a pulse scheme with a total duration of $2912,dt$ is required.}
    \label{fig: transpile_IBM_sherbrook_circuit_Bell}.
\end{figure}

\begin{figure}[H]
    \centering
    \includegraphics[width=0.7\linewidth]
    {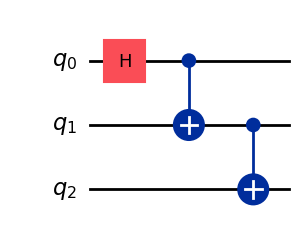}
\caption{Representation of the GHZ state $|GHZ\rangle = \frac{1}{\sqrt{2}}\left(|000\rangle +|111\rangle \right) $ in a quantum circuit diagram. The sate is obtained by implementing a Hadamard gate on $q_{0}$ and two consecutive CNOTs gates on qubits $q_{0}$ to $q_{1}$ and $q_{1}$ to $q_{2}$ respectively.}
    \label{fig: circuit_GHZ_state}
\end{figure}

\begin{figure}[H]
    \centering
    \includegraphics[width=1\linewidth]
    {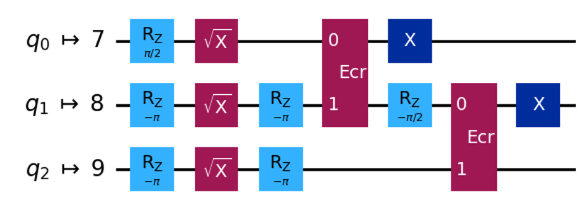}
\caption{Representation of the circuit shown in Fig.~\ref{fig: circuit_GHZ_state} using the native gates of the IBM$\_$sherbrooke. This circuit contains two bipartite gates. To reproduce this state, the corresponding pulse sequence has a total duration of $5315,dt$}
    \label{fig:transpile_IBM_sherbrook_circuit_GHZ}
\end{figure}

\begin{figure}[H]
    \centering
    \includegraphics[width=1\linewidth]
    {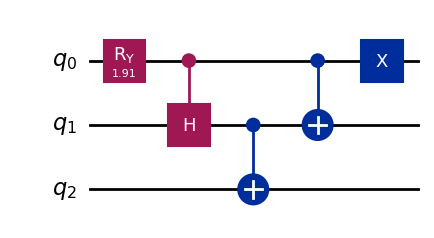}
\caption{Representation of the state $|W\rangle = \frac{1}{\sqrt{3}}\left(|100\rangle + |010\rangle + |001\rangle\right)$ using a quantum circuit scheme. In this case, an  $R_{Y}$ gate, a Hadamard gate, two CNOT gates, and an $X$ gate are represented.}
    \label{fig:circuit_W_state}
\end{figure}

\begin{figure}[H]
    \centering
    \includegraphics[width=1\linewidth]
    {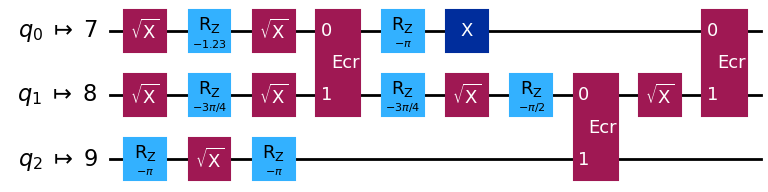}
\caption{Representation of the scheme of fig. \ref{fig:circuit_W_state} in the native gates of the IBM$\_$sherbrooke simulator,It can be noted that there are three interleaved bipartite gates. The duration of this circuit is $8224,dt$.}
    \label{fig:transpile_IBM_sherbrook_circuit_w}
\end{figure}

Each gate represents a pulse sequence that acts on some of the qubits, changing their energy states through their interaction with the pulses. Local gates are implemented with a single pulse, while bipartite gates, such as the ECR, are implemented using a sequence of up to five pulses. In this research, the implemented pulse sequences are based on the transpiled circuits shown in the figures.\ref{fig: transpile_IBM_sherbrook_circuit_Bell}, \ref{fig:transpile_IBM_sherbrook_circuit_GHZ} and \ref{fig:transpile_IBM_sherbrook_circuit_w}.
\\\\
It is worth noting that the times required to reproduce the pulse sequences obtained for the schemes studied in this work are shorter than those given by default in the simulator, as shown in \ref{fig: state_bell2_gaussian}, \ref{fig: pulse_ghz_gaussian}, and \ref{fig: pulse_w_gaussian}. This observation is particularly relevant in the context of this work, in which decoherence was not considered. However, the obtained parameters can serve as a reference for considering this approach in realistic systems, since some qubits may exhibit coherence times longer than the duration of the proposed pulse sequences.

\section{Electromagnetic pulse modeling and envelope functions}
\label{APPENDIX B}

The quantum circuits used to generate the states of interest in this research, as well as their decomposition into the native gates of a quantum computer, have been presented. However, in a physical implementation, these gates are realized through electromagnetic pulses. These pulses are characterized by a frequency and are modulated by a function defined by a set of parameters.
\\\\
 In the control Hamiltonians given by Eqs \ref{eq_3}, \ref{eq_6}, and \ref{eq_8}, the function $\mu_{i}(t)$ is considered the envelope function, which must satisfy certain restrictions. Furthermore, the implementation of this function depends on the waveform generator used in the quantum computer hardware \cite{matekole2022methods}. For example, in the simulator and on the IBM\_Sherbrooke computer, Gaussian, Gaussian Square, DRAG, and constant functions can be implemented \cite{alexander2020qiskit}.

\begin{figure}[H]
    \centering
    \includegraphics[width=1\linewidth]
    {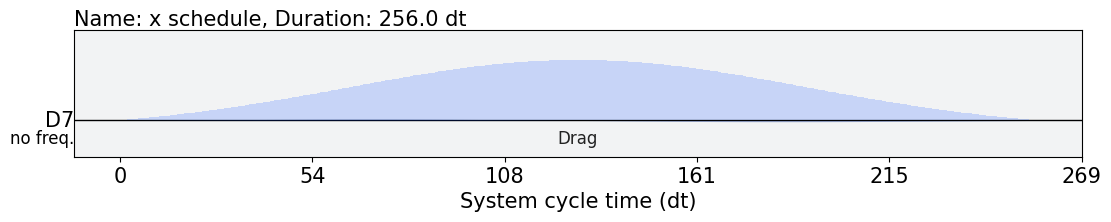}
\caption{The envelope function used to implement the quantum gate $X$ on the IBM\_Sherbrooke computer is known as DRAG and consists of five parameters.}
    \label{fig: x_schedule}
\end{figure}

\begin{figure}[H]
    \centering
    \includegraphics[width=1\linewidth]
    {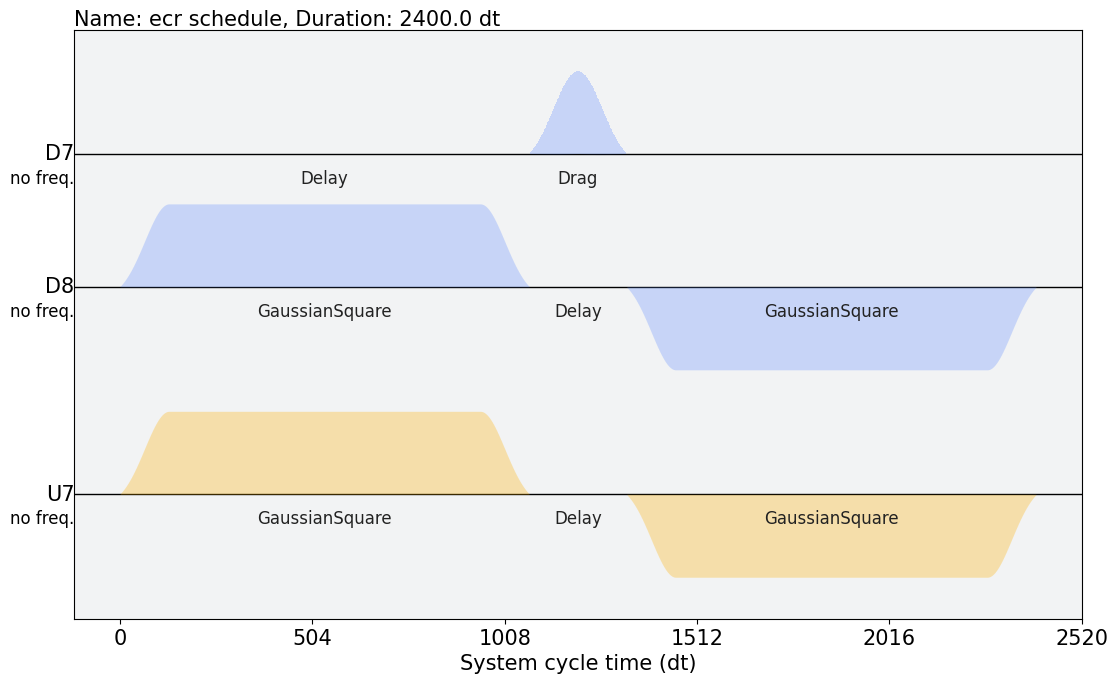}
\caption{Pulse scheme for reproducing the ECR quantum gate of the IBM\_Sherbrooke computer. The sequence consists of five pulses, where qubit seven is the control qubit and qubit eight is the target qubit. The implementation requires a DRAG function (see top line) and four Gaussian Square functions.}
    \label{fig: ecr schedule}
\end{figure}

The preceding figures illustrate how quantum gates are implemented using pulse schemes. These functions are modulated by a cosine function whose frequency must match that of the qubit being controlled. By controlling the parameters of the envelope functions, the desired quantum gates can be constructed. However, for bipartite gates, additional pulses are required because the interaction between the qubits is more complex.
\\
For our implementation, we first used constant pulses to evaluate the desired states and then employed more realistic pulses, such as Gaussian Square pulses, since they behave similarly to constant-amplitude pulses.\cite{zhu2025hardware}.

\subsection{Gaussian Square}

The Gaussian Square pulse is a control signal used in superconducting qubits that combines a central region of approximately constant amplitude with smooth Gaussian-type transitions at its ends. Its main parameters are the amplitude ($A$), total duration ($T$), and width of the central region ($W$), which determines the duration of the constant-amplitude region. The signal can also incorporate a phase $(\phi)$ and a carrier frequency $(\omega_{d})$. This pulse shape is used in quantum control because it allows for controlled rotations of the qubit state while providing smooth transitions in the signal, thereby reducing the excitation of higher transmon levels and improving the fidelity of quantum operations.\cite{gambetta2011analytic}.

Mathematically, Qiskit defines a Gaussian Square pulse as

% \begin{equation}
%  f'(x) = \left\{  
% \begin{array}{ll}
%     \exp\left(-\frac{1}{2} \frac{(x - risefall)^{2}}{\sigma^{2}}\right) & x < risefall \\
%      1                  & risefall \leq x  < risefall + width  \\
%      \exp\left(-\frac{1}{2} \frac{(x - (risefall + width)^{2}}{\sigma^{2}}\right) & risefall + width \geq x
% \end{array}
% \right.
% \end{equation}

\begin{equation}
\resizebox{\columnwidth}{!}{$
f'(x)=
\begin{cases}
\exp\left[-\dfrac{(x-risefall)^2}{2\sigma^2}\right],
& x<risefall,\\
1,
& risefall\leq x<risefall+width,\\
\exp\left[-\dfrac{(x-risefall-width)^2}{2\sigma^2}\right],
& x\geq risefall+width.
\end{cases}
$}
\end{equation}

\begin{equation}
\begin{array}{cc}
   f(x) = A\frac{f'(x)-f(1)}{1-f'(1)},   &  0\leq x < duration    
\end{array}
\end{equation}

where $f'(x)$ is the gaussian square waveform without  lifting of  amplitude, and $A = amp \times\exp{(i\times angle)}$ 

If $risefall\_sigma\_ratio$ is not None and width is None:
$$risefall = risefall\_sigma\_ratio \times \sigma $$
$$width = duration - 2 \times risefall$$

If width is not None and $ risefall\_sigma\_ratio$ is None:

$$ risefall = \frac{duration -width}{2} $$

Parameters

\begin{itemize}
    \item Duration - pulse length in terms of the sampling period $dt$.

    \item amp - The magnitude of the amplitude of the Gaussian and square pulse.

    \item $\sigma$ - A measure of how wide or narrow the Gaussian risefall is; see the class docstring  for more details.

    \item width - The duration of the embedded square pulse.

    \item angle - The angle of the complex amplitude of the pulse. Default value 0.

    \item $risefall\_sigma\_ratio$ - The ratio of each risefall duration to $\sigma$.
\end{itemize}

Table \ref{tab: Gaussian pulse data} shows the parameters obtained for each of the entanglement class simulations, where $t_{i}$ refers to the duration. Values of 0 and 1 were used for the angle and $risefall\_sigma\_ratio$, respectively.

\begin{figure}[H]
    \centering
    \includegraphics[width=1\linewidth]{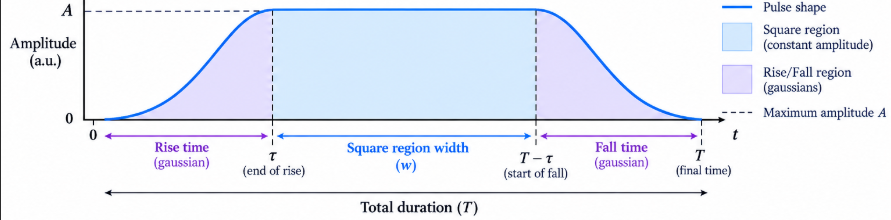}
    \caption{Representation of Gaussian pulse parameters}
    \label{fig:Gaussian representation parameters}
\end{figure}

%%%%%%%%%%%%%%%%%%%%%%%%%%%%%%%%%%%%%%%%%%%%%%%%%%

\section{Numerical data}
\label{Appendix C}

In this section, we present the numerical results obtained for the reconstruction of the states of interest, along with the Bures distance and other entanglement measures.

\subsection{Amplitude and time data for square pulses}

Here, we present the square-pulse parameters used to reproduce the states described by Eqs.~\ref{distri dos qubits}, \ref{GHZ-cuadrado}, and \ref{W-cuadrado}. These parameters were obtained from numerical simulations by solving the Schr\"odinger equation.

\begin{table}[H] % O [htbp]
\centering % Comando para centrar la tabla
\begin{tabular}{c|c|c|c} 
Pulse & $Bell$ & $GHZ$ & $W$    \\ 
\hline 
$Amp_{1}$&0.6824432988764655    & 0.490394435      & -0.204862473\\ 
$t_{1}$ &191.910934             & 191.779586       &  234.830783 \\ 
$Amp_{2}$ &0.8286644926788276   &-0.691612520      & 0.253423969\\
$t_{2}$ &191.910934             &  191.779586      & 234.830783\\
$Amp_{3}$&0.7868326972398657    & 0.665492967      & -0.675297508\\
$t_{3}$ &797.242687             &1037.50354        & 234.830783\\
$Amp_{4}$ &                     &  0.165200193     &-0.511593803\\
$t_{4}$ &                       &168.736170        & 1565.74174\\
$Amp_{5}$ &                     & 0.19735822       &-0.236750884\\
$t_{5}$ &                       &1450.46248   & 187.479735\\
$Amp_{6}$ &                     &             & -0.853527417\\
$t_{6}$ &                       &             & 187.479735\\
$Amp_{7}$ &                     &             & 0.02335491\\
$t_{7}$ &                       &                 &187.479735\\
$Amp_{8}$ &                     &                 & -0.853527417\\
$t_{8}$ &                       &                 & 1229.42236\\
$Amp_{9}$ &                     &                 & 0.02335491\\
$t_{9}$ &                       &                 &274.196599\\
$Amp_{10}$ &                    &                 &-0.444121784\\
$t_{10}$ &                      &                 &274.196599\\
$Amp_{11}$ &                    &                 &0.866030588\\
$t_{11}$ &                      &                  &274.196599\\
$Amp_{12}$ &                    &                 & -0.605000131\\
$t_{12}$ &                      &                 &1323.52652\\

\end{tabular} 
\caption{Data for the parameters obtained from the simulation for square pulses. The table shows the parameters for the schemes \ref{fig: pulse_bell2_square},\ref{fig: pulse_ghz_square},\ref{fig: pulse_w_square}. Some amplitudes are negative because a phase of $180^\circ$ is assumed. Here, $t_n$ refers to the pulse times in $ns$, and $Amp_n$ refers to the amplitude.} 
\label{tab: datos pulsos cuadrados} 
\end{table}

\subsection{Bures Distance}
\label{apen Bures Distance}

The Bures distance is a measure of distance between quantum states, defined on the space of density matrices and closely related to quantum fidelity and the geometry of the state space \cite{Bures1969,Uhlmann1976}. For two quantum states $\rho$ and $\sigma$, the Bures distance is defined as

\begin{equation}
    D_{\mathrm{B}}(\rho,\sigma) = \sqrt{2\left[1-\sqrt{F(\rho,\sigma)}\right]}
\end{equation}

where $F(\rho,\sigma)$ corresponds to the Uhlmann fidelity, given by

\begin{equation}
    F(\rho,\sigma) = \left[\operatorname{Tr} \sqrt{\sqrt{\rho},\sigma,\sqrt{\rho}} \right]^{2}
\end{equation}

The Bures distance satisfies $D_{\mathrm{B}}(\rho,\sigma)=0$ if and only if $\rho = \sigma$, while it reaches its maximum value $\sqrt{2}$ or mutually orthogonal states. In the particular case of pure states, $\rho=|\psi\rangle\langle\psi|$ and $\sigma=|\phi\rangle\langle\phi|$, this expression reduces to

\begin{equation}
    D_{\mathrm{B}}(\psi,\rho) = \sqrt{2\left(1-|\langle\psi|\phi\rangle|\right)}
    \label{Distance-Bures}
\end{equation}

n order to determine the local unitary transformations that approximate the state obtained by simulating quantum dynamics to the objective states $|W\rangle$ and $|GHZ\rangle$, a global transformation of the form $(U_{\mathrm{L}}=U_1\otimes U_2\otimes U_3)$, where each  $(U_i)$ corresponds to a single unitary operation parametrized by three angles. Starting from the simulated state $( |\psi_{\mathrm{sim}}\rangle )$, the transformed state $( |\psi_{\mathrm{L}}\rangle=U_{\mathrm{L}}|\psi_{\mathrm{sim}}\rangle )$ iis constructed, and the nine parameters of the local operations are determined by a numerical optimization procedure that seeks to minimize the Bures distance \eqref{Distance-Bures}. In this way, the optimal set of parameters corresponds to the local transformations that produce states closest to the $W$ and $GHZ$ states, so that a value of $D_{\mathrm{B}}$ close to zero indicates a high degree of equivalence between the two states. This methodology allows us to numerically establish whether the state generated by the dynamics can be transformed into the $W$ and $GHZ$ states through local unitary operations, thus providing a criterion for analyzing their equivalence under local transformations.

\begin{center}
\begin{table}[H]
\begin{tabular}{c|c|c|c|c|c|c|}
unitary operator & $\theta_{1}$ & $\theta_{2}$ & $\theta_{3}$ & Bures distance & Fidelity \\
\hline
$U_{1}$ & 4.120    &  5.077   & 3.926 &         &         \\
$U_{2}$ & 0.4017   & 0.001479 & 4.909 &  0.056  &   0.996  \\  
$U_{3}$ & 1.876    & 2.243    & 6.014 &         &           \\
\end{tabular}
\caption{Values of the parameters of each unitary operation used to obtain the Bures distance closest to the $W$ state. Here, $\theta$ denotes the parameters of each unitary operation, measured in radians. The calculated Bures distance and fidelity are also shown.}
\end{table}
\end{center}

\begin{center}
\begin{table}[H]
\begin{tabular}{c|c|c|c|c|c|c|}
unitary operator & $\theta_{1}$ & $\theta_{2}$ & $\theta_{3}$ & Bures distance & Fidelity \\
\hline
$U_{1}$ & 2.698    &  1.284  & 5.239 &         &         \\
$U_{2}$ & 2.788    &4.915    & 3.721 &  0.011  &   0.999  \\  
$U_{3}$ & 2.767    &5.063    & 2.120 &         &           \\
\end{tabular}
\caption{Values of the parameters of each unitary operation used to obtain the Bures distance closest to the $GHZ$ state. Here, $\theta$ denotes the parameters of each unitary operation, measured in radians. The calculated Bures distance and fidelity are also shown.}
\end{table}
\end{center}

\subsection{Gaussian pulse parameters}

To obtain the states \ref{psi_gauss}, \ref{GHZ_gauss}, and \ref{eq: gauss_w}, the simulation was performed using Gaussian pulses, for which the following parameters were obtained. It should be noted that the pulse shape is a Gaussian square; for more information, see Appendix \ref{APPENDIX B}.

\begin{table}[H] % O [htbp]
\centering % Comando para centrar la tabla
\begin{tabular}{c|c|c|c} 
Pulse & $Bell$ & $GHZ$ & $W$    \\ 
\hline 
$Amp_{1}$&0.6824432988764655       &0.80939802     &-0.499757010                            \\ 
$t_{1}$ &345.4                     & 413.910934    & 360.830783                           \\ 
$\sigma_{1}$&72.18                 &111.40668999   & 63.7831081                            \\
$width_{1}$&  191.910934           &191.910934     & 234.830783                           \\
$Amp_{2}$&0.8286644926788276       &0.27798819     &-0.155218319                            \\
$t_{2}$  & 345.4                   &413.910934     & 360.830783                            \\
$\sigma_{2}$ & 72.18               &111.40668999   &63.7831081                             \\
$width_{2}$ & 191.910934           &191.910934     &234.830783                             \\
$Amp_{3}$ &0.7868326972398657      &0.69514675     &-0.698213190                        \\
$t_{3}$   &1034                    &1297.50354     & 360.830783                        \\
$\sigma_{3}$& 97.56                &130.31892528   &63.7831081                          \\
$width_{3}$ & 797.242687           &103.750354     &234.830783                          \\
$Amp_{4}$   &                      &0.81862309     &0.224520443                             \\
 $t_{4}$    &                      &304.73617      &1607.74174                       \\
 $\sigma_{4}$&                     &68.31986628    &121.759010                        \\
 $width_{4}$&                      &168.736170     &1565.74174                        \\
 $Amp_{5}$  &                      &0.23133505       &0.612171915                           \\
 $t_{5}$     &                     &1734.46248       &553.479735                           \\
 $\sigma_{5}$&                     &142.62017516     &183.590195                             \\
 $width_{5}$&                      &145.046248       &187.479735                          \\
$Amp_{6}$  &                        &                 &0.0256210429                           \\
 $t_{6}$     &                      &                 &553.479735                           \\
 $\sigma_{6}$&                      &                 &183.590195                             \\
 $width_{6}$&                       &                 &187.479735                           \\
 $Amp_{7}$  &                       &                 & -0.0938807900                          \\
 $t_{7}$     &                      &                 &553.479735                           \\
 $\sigma_{7}$&                      &                 &183.590195                             \\
 $width_{7}$&                       &                 &187.479735                            \\
 $Amp_{8}$  &                       &                 &-0.798677953                           \\
 $t_{8}$     &                      &                 &1405.42236                           \\
 $\sigma_{8}$&                      &                 &88.0125865                           \\
 $width_{8}$&                       &                 &1229.42236                           \\
 $Amp_{9}$  &                       &                 &-0.0606957067                           \\
 $t_{9}$     &                      &                 &300.196599                              \\
 $\sigma_{9}$&                      &                 &113.576045                           \\
 $width_{9}$&                       &                 &274.196599                           \\
 $Amp_{10}$  &                      &                 &-0.902406925                           \\
 $t_{10}$     &                     &                 &300.196599                           \\
 $\sigma_{10}$&                     &                 &113.576045                           \\
 $width_{10}$&                      &                 &274.196599                          \\
 $Amp_{11}$  &                      &                 &0.51397687                           \\
 $t_{11}$     &                     &                 &300.196599                           \\
 $\sigma_{11}$&                     &                 &113.576045                           \\
 $width_{11}$&                      &                 &274.196599                          \\
 $Amp_{12}$  &                      &                 &-0.6666759                           \\
 $t_{12}$     &                     &                 &1507.52652                          \\
 $\sigma_{12}$&                     &                 & 92.9872143                           \\
 $width_{12}$&                      &                 & 1323.52652                          \\

\end{tabular} 
\caption{Parameter data obtained from the simulation for Gaussian square pulses. The table shows the parameters for the schemes \ref{fig: state_bell2_gaussian},\ref{fig: pulse_ghz_gaussian},\ref{fig: pulse_w_gaussian}. Some amplitudes are negative because a phase of $180^\circ$ is assumed. The other values are given in $ns$.}
\label{tab: Gaussian pulse data}
\end{table}
States obtained for square Gaussian pulses

\subsection{States obtained for square Gaussian pulses}

The state was obtained from the simulation of the scheme \ref{fig: pulse_bell2_gaussian} using Gaussian pulses. The simulation was performed for two systems with three energy levels.

\begin{equation}
\begin{split} 
|\Psi_{gauss} \rangle  =     & (0.44758 + 0.27527) |00\rangle \\
&+ (0.10419  -0.46041i) |01\rangle  \\
&+ (-0.00067 -0.00221i)|02\rangle  \\
&+(-0.41510  -0.24113i) |10\rangle  \\
&+( 0.06061  -0.51645i) |11\rangle   \\
&+ (-0.00219 + 0.00495i) |12\rangle  \\
&+ (0.00723  -0.00396i) |20\rangle  \\
&+ (0.00309  +0.00513i) |21\rangle  \\
&+ (-0.00000 -0.00003i) |22\rangle  \\
    \end{split}
    \label{psi_gauss}
\end{equation}

The state was obtained from the simulation of the scheme \ref{fig: pulse_ghz_gaussian} with Gaussian pulses; the simulation was done for three systems with three energy levels.

\begin{equation}
\begin{split}
|\Psi'_{\mathrm{GHZ\_gauss}}\rangle
=\, &+(0.05489+0.06235\,i)\,|000\rangle\\
&+(0.19308-0.53558\,i)\,|001\rangle\\
&+(0.00195-0.00175\,i)\,|002\rangle\\
&+(-0.31596-0.25770\,i)\,|010\rangle\\
&+(0.00942-0.05850\,i)\,|011\rangle\\
&+(0.00269-0.00199\,i)\,|012\rangle\\
&+(0.00327+0.01199\,i)\,|020\rangle\\
&+(0.00677-0.00043\,i)\,|021\rangle\\
&+(-0.00000-0.00000\,i)\,|022\rangle\\
&+(-0.01049+0.05800\,i)\,|100\rangle\\
&+(0.38965-0.11660\,i)\,|101\rangle\\
&+(0.00101+0.00064\,i)\,|102\rangle\\
&+(-0.03579+0.56721\,i)\,|110\rangle\\
&+(-0.07706+0.03301\,i)\,|111\rangle\\
&+(-0.00034-0.00072\,i)\,|112\rangle\\
&+(0.00284-0.00380\,i)\,|120\rangle\\
&+(0.00441-0.00088\,i)\,|121\rangle\\
&+(0.00001+0.00001\,i)\,|122\rangle\\
&+(-0.00049+0.00369\,i)\,|200\rangle\\
&+(-0.00104+0.00215\,i)\,|201\rangle\\
&+(-0.00001-0.00000\,i)\,|202\rangle\\
&+(-0.00073-0.00195\,i)\,|210\rangle\\
&+(0.00029-0.00026\,i)\,|211\rangle\\
&+(-0.00000+0.00001\,i)\,|212\rangle\\
&+(-0.00001+0.00002\,i)\,|220\rangle\\
&+(-0.00002+0.00002\,i)\,|221\rangle\\
&+(-0.00001-0.00000\,i)\,|222\rangle.
\end{split}
\label{GHZ_gauss}
\end{equation}

The state was obtained from the simulation of the scheme \ref{fig: pulse_w_gaussian} using Gaussian pulses. The simulation was performed for three systems with three energy levels.

\begin{equation}
\begin{split}
|\Psi'_{w\_gauss}\rangle
=\, &(0.06927+0.01477\,i)\,|000\rangle\\
&+(0.42353-0.02411\,i)\,|001\rangle\\
&+(0.08602+0.00579\,i)\,|002\rangle\\
&+(0.24681-0.09064\,i)\,|010\rangle\\
&+(-0.39360+0.29951\,i)\,|011\rangle\\
&+(0.00128-0.05533\,i)\,|012\rangle\\
&+(-0.04128+0.03876\,i)\,|020\rangle\\
&+(0.21157-0.00836\,i)\,|021\rangle\\
&+(-0.00297+0.00162\,i)\,|022\rangle\\
&+(-0.01831-0.05047\,i)\,|100\rangle\\
&+(-0.17494-0.31504\,i)\,|101\rangle\\
&+(-0.01218-0.06149\,i)\,|102\rangle\\
&+(-0.31455-0.35828\,i)\,|110\rangle\\
&+(-0.23700-0.07802\,i)\,|111\rangle\\
&+(-0.02685+0.03187\,i)\,|112\rangle\\
&+(0.09112-0.02269\,i)\,|120\rangle\\
&+(-0.07631-0.00933\,i)\,|121\rangle\\
&+(0.00092+0.00036\,i)\,|122\rangle\\
&+(-0.00241-0.00207\,i)\,|200\rangle\\
&+(0.00010-0.00059\,i)\,|201\rangle\\
&+(-0.00018+0.00032\,i)\,|202\rangle\\
&+(0.00362+0.00057\,i)\,|210\rangle\\
&+(-0.00139-0.00141\,i)\,|211\rangle\\
&+(0.00007+0.00014\,i)\,|212\rangle\\
&+(0.00006-0.00027\,i)\,|220\rangle\\
&+(-0.00022+0.00049\,i)\,|221\rangle\\
&+(-0.00001+0.00001\,i)\,|222\rangle.
\end{split}
\label{eq: gauss_w}
\end{equation}

\section{Numerical data for the closed system in two qubits}
\label{Numerical data for the open system in two qubits}

To characterize the oscillations observed in the graph of the closed system, whose behavior coincides with that of the open system, after the application of the resonant pulse between the qubits, a signal interval was initially selected. Specifically, a segment was considered that spanned from the instant immediately after the application of the control pulse (excluding the pulses from channels D0 and D1) to a later moment when this oscillatory behavior was still present. This interval was used as a reference to characterize and determine the oscillations present in the system's dynamics.

To quantitatively characterize the observed dynamics, the data were fitted using the function $ a + bt + A_{1}cos(\omega_{1}t + \phi_{1}) + A_{2}cos(\omega_{2}t + \phi_{2}) + A_{3}cos(\omega_{3}t + \phi_{3})$. This The fitting function combines a linear trend with three oscillatory contributions of different frequencies. The term $(a+bt)$ describes the overall evolution of the signal, while each oscillatory component is characterized by its amplitude $(A_i)$, angular frequency $(\omega_i)$, and initial phase $(\phi_i)$, with $(i=1,2,3)$. This parameterization allows the identification and quantification of the different time scales present in the dynamics, providing a simultaneous description of the background trend and the observed oscillations. The parameter values obtained from the fit are presented in Table \ref{tab: ajustes}.

\begin{figure}[H]
    \centering
    \includegraphics[width= 1\linewidth]{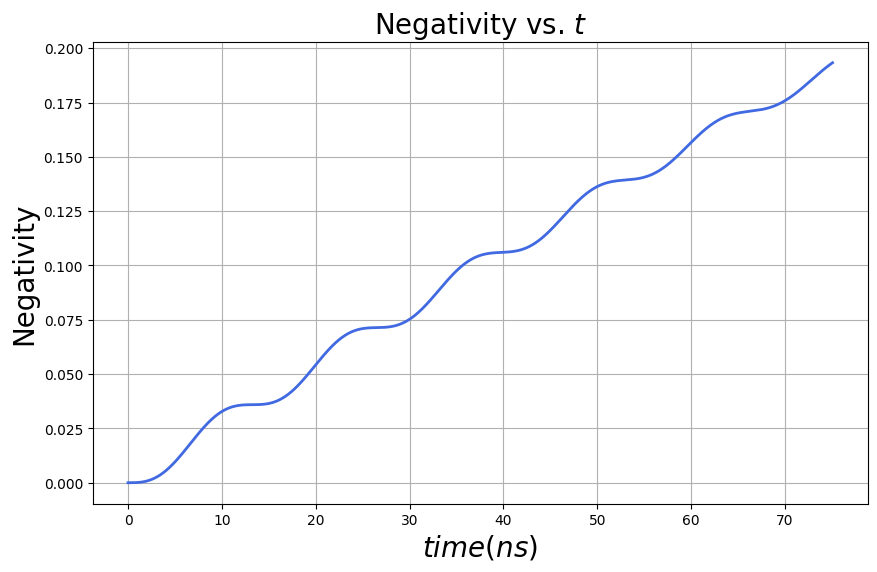}
    \caption{Negativity as a function of time for the closed system, considering the application of the control pulse on qubit 8. The oscillations present in the dynamics of the negativity are observed.}
    \label{tramo-open-2qubits}
\end{figure}

\begin{table}[H]
    \centering
    \begin{tabular}{c|c|c}
       Quantity  & & Fitted value  \\
       \hline
          A1                 &    &  -5.92955368e-01 ± 2.35e-02          \\
          $\omega_{1}$       &    &   6.76673507e-03 ± 9.16e-05 $rad/ns$         \\
          $\phi$             &    &  1.57424241e+00 ± 1.13e-04          \\   
          $f_{1}$            &    &   1.07695934 ± 0.01457796   MHz      \\
          $T_{1}$            &    &   928.54016566 ± 12.56892259 ns        \\
          A2                 &    &  -2.98701772e-03 ± 8.53e-07          \\
          $\omega_{2}$       &    &  4.61006934e-01 ± 5.81e-06  $rad/ns$        \\
          $\phi_{2}$         &    &  -1.55003678e+00 ± 1.93e-04         \\   
          $f_2$              &    &  73.37153234 ± 0.00092412 MHz         \\
          $T_2$              &    &   13.62926421 ± 0.00017166 ns        \\
           A3                &    &  2.72899456e-03 ± 8.32e-07          \\
          $\omega_{3}$       &    &  4.87449974e-01 ± 5.73e-06 $rad/ns$           \\
          $\phi_{3}$         &    &  1.54823399e+00 ± 2.05e-04          \\   
          $f_3$              &    &   77.58007285 ± 0.00091214 MHz        \\
          $T_3$              &    &    12.88990798 ± 0.00015155 ns        \\
           $R^{2}$           &    &    1.00000000       \\
           RMSE              &    &    3.02811625e-06       \\
    \end{tabular}
    \caption{The table reports the fitted parameters together with their standard uncertainties, as well as the corresponding oscillation frequency $(f)$, period $(T)$, and the goodness-of-fit indicators $(R^{2})$ and RMSE.}
    \label{tab: ajustes}
\end{table}

\begin{figure}[H]
    \centering
    \includegraphics[width= 1\linewidth]{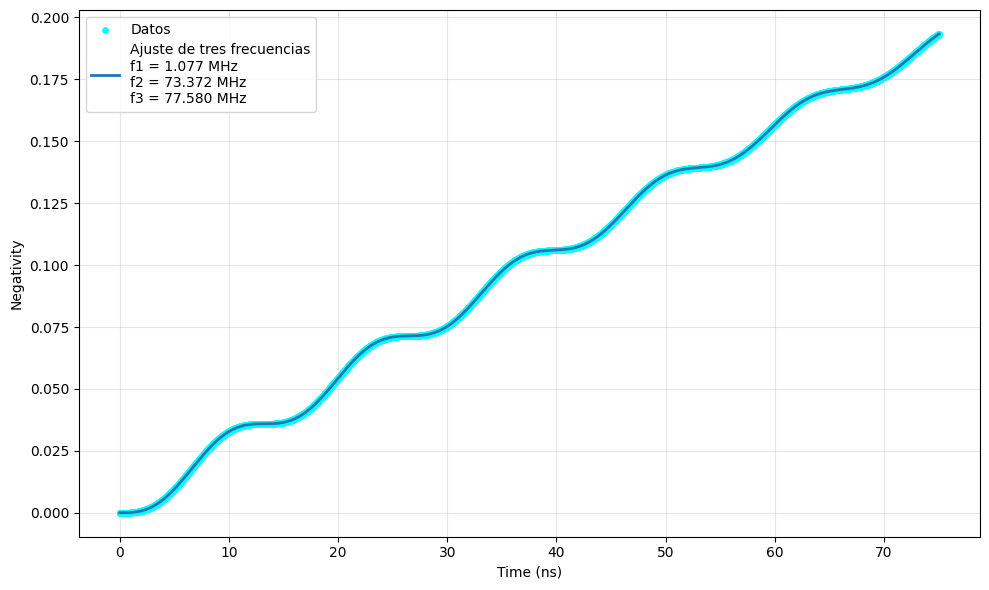}
    \caption{Time-dependent negativity for the coupled 2-qubit system, considering the application of the control pulse only to the closed system. The cyan points correspond to the simulated data, and the blue curve corresponds to the nonlinear least-squares fit with the model. $ a + bt + A_{1}cos(\omega_{1}t + \phi_{1}) + A_{2}cos(\omega_{2}t + \phi_{2}) + A_{3}cos(\omega_{3}t + \phi_{3})$, which captures the increasing trend of negativity superimposed on periodic oscillations.}
    \label{fig:ajuate_2}
\end{figure}

\begin{table}[H]
\centering
\begin{tabular}{c|c|c|c|c|c|}
frequencies (MHz) & $\delta = |\omega_{7} -\omega_{8}|/2\pi$ & $J_{78}/2\pi$ & $r_{0}/2\pi$ & $r_{1}/2\pi$  \\
\hline
1.0769  &          &          &          &                 \\
73.3715 &  56.9437 & 2.013    &  62.9868 &     103.4740         \\  
77.5800 &          &          &          &                  \\
\end{tabular}
\caption{Comparison between the frequencies obtained from the negativity dynamics fitting and the characteristic frequency scales of the two-qubit Hamiltonian. The qubit frequency mismatch is given by $\delta=|\omega_7-\omega_8|/2\pi$, while the coupling strength $J_{78}/2\pi$ and the excitation amplitudes $r_0/2\pi$ and $r_1/2\pi$ are expressed in MHz.}
\label{tab: ajuste -frecue}
\end{table}

\begin{figure}[H]
    \centering
    \includegraphics[width=1\linewidth]{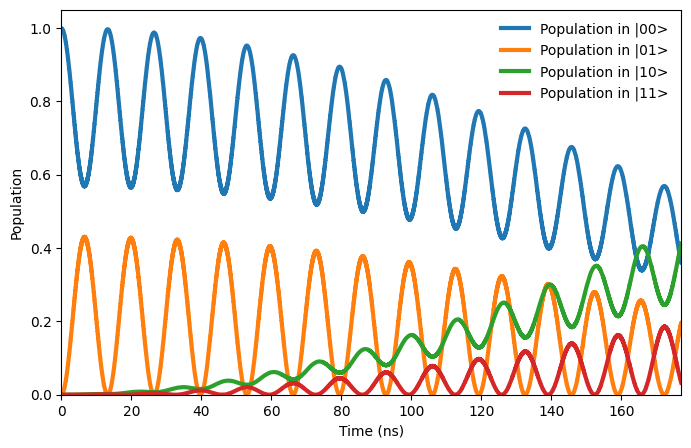}
    \caption{Time evolution of the populations of the four states of the computational basis $(|00\rangle, |01\rangle, |10\rangle, |11\rangle)$ for the closed system of two coupled qubits, under the exclusive application of the control pulse on qubit 8, without including the additional local pulses. The system starts from the initial state $|00\rangle$, and an oscillatory population transfer is observed predominantly toward the state $|01\rangle$, with a decreasing envelope in the amplitude of the oscillations of $|00\rangle$ and $|01\rangle$ as time passes. Simultaneously, $|01\rangle$ and $|11\rangle$ gradually grow, demonstrating the progressive emergence of dynamics in the two-excitation subspace, attributable to the exchange coupling $J_{78}$ between the qubits. This behavior reflects the nontrivial dynamics induced by the control pulse in the absence of compensation by local pulses. }
    \label{fig:placeholder}
\end{figure}

Table \ref{tab: ajuste -frecue} shows that the dynamic oscillation of the negativity can be described by three frequency components: $f_1=1.07696$ MHz, $f_2=73.37153$ MHz, and $f_3=77.58007$ MHz. These frequencies are compared to the main dynamic scales present in the Hamiltonian, given by $\delta/2\pi=56.9437$ MHz, $J_{78}/2\pi=2.0136$ MHz, $r_{0}/2\pi=62.9868$ MHz, and $r_{1}/2\pi=103.4740$ MHz.

It is important to note that the system is in heavily detuned regime, since that

$$\frac{\delta}{J_{78}} \approx 28.3$$

Therefore, the coupling between the qubits is considerably smaller than the difference between their transition frequencies, and the entanglement observed during the evolution should not be interpreted as a consequence of a direct resonant exchange of excitations between the qubits.

The higher frequency components, $f_{2}$ and $f_{3}$, are in a region comparable to the scales associated with detuning and the drive terms. However, this numerical proximity does not allow for a direct correspondence between each fitted frequency and an individual parameter of the Hamiltonian. The frequencies extracted from the fitting should be interpreted as effective dynamical frequencies, which result from the joint evolution of the system under the simultaneous action of the coupling between qubits and the control fields.

Furthermore, the separation between components $f_{2}$ and $f_{3}$ produces a beat modulation characterized by $ f_{\mathrm{beat}} = |f_{3} - f_{2}| = 4.20854  $  MHz

This value is numerically close to $\frac{2J_{78}}{2\pi} = 4.0272$ MHz. However, because the system is heavily detuned and simultaneously subjected to drive fields, this coincidence is not sufficient to attribute the origin of the beat exclusively to the coupling $J_{78}$. Consequently, $f_{\mathrm{beat}}$ must be considered an effective frequency associated with the combined dynamics of the system.

Taken together, the results indicate that the observed oscillations in the negativity are not determined by a single frequency scale, but rather emerge from the interaction between detuning, the fixed coupling between the qubits, and the applied control fields. The low-frequency component $(f_1)$, whose amplitude is dominant, describes the main time evolution of the negativity, while the components $(f_2)$ and $(f_3)$ introduce higher-frequency modulations that contribute to the observed oscillatory structure.

%%%%%%%%%%%%%%%%%%%%%%%%%%%%%%%%%%%%%%%%

\section{Comparison of different simulations by varying the Hamiltonian parameters}

This section presents the results of simulations corresponding to different variations of the Hamiltonian. The analysis was conducted on an isolated system of two superconducting qubits \ref{sub: two qubits}, employing the square-pulse scheme designed to generate a Bell state \ref{fig: pulse_bell2_square}. To evaluate the influence of the various Hamiltonian terms, simulations were performed by controllably modifying their composition and comparing the resulting entanglement dynamics.

\begin{figure}[H]
    \centering
    \includegraphics[width=1.0\linewidth]{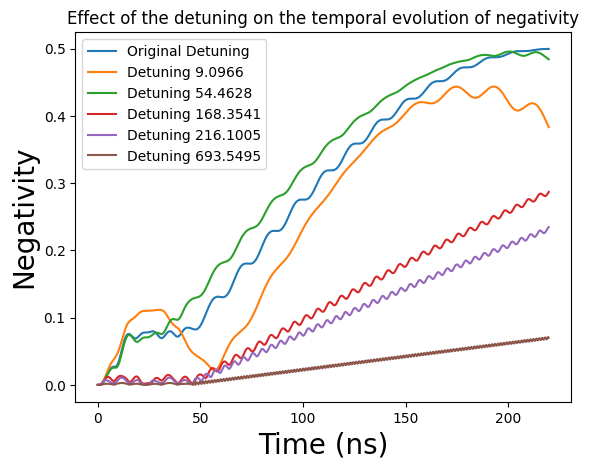}
    \caption{Comparison of the negativity dynamics for different detuning values, expressed in MHz. The blue curve corresponds to the Hamiltonian with the original detuning.}
    \label{nega-detunig}
\end{figure}

The Fig \ref{nega-detunig} shows that detuning significantly influences the generation and temporal evolution of entanglement. For values of $\lvert\delta\rvert$ greater than the coupling strength, the negativity remains low during the application of pulses $D0$ and $D1$ but increases upon application of the cross-resonance pulse. Furthermore, as $\lvert\delta\rvert$ increases, oscillations in the negativity become faster and exhibit reduced amplitude. In contrast, small detuning values—or those comparable to the coupling strength—favor more pronounced entanglement dynamics.

\begin{figure}[H]
    \centering
    \includegraphics[width=1.0\linewidth]{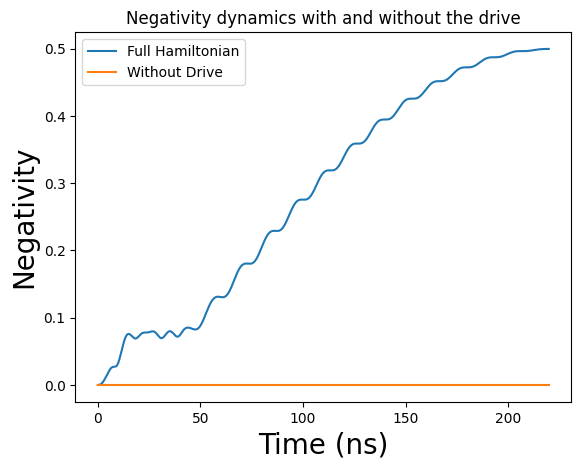}
    \caption{Comparison of the negativity dynamics with and without external driving. The blue curve corresponds to the full Hamiltonian, while the orange curve is obtained by removing the driving term. The absence of entanglement generation without driving highlights the role of the external fields in creating the excitations required for the coupling interaction to generate entanglement.}
    \label{nega-sin-drives}
\end{figure}

Figure. \ref{nega-sin-drives} compares the time evolution of the negativity for the full Hamiltonian—including control pulses—with that of the case without external excitations. In the absence of pulses, the negativity remains virtually zero throughout the evolution, indicating that, under the conditions considered, qubit coupling alone does not produce an appreciable increase in entanglement. Entanglement generation therefore requires the combined action of the coupling and the external excitations introduced via the control pulses.

\begin{figure}[H]
    \centering
    \includegraphics[width=1.0\linewidth]{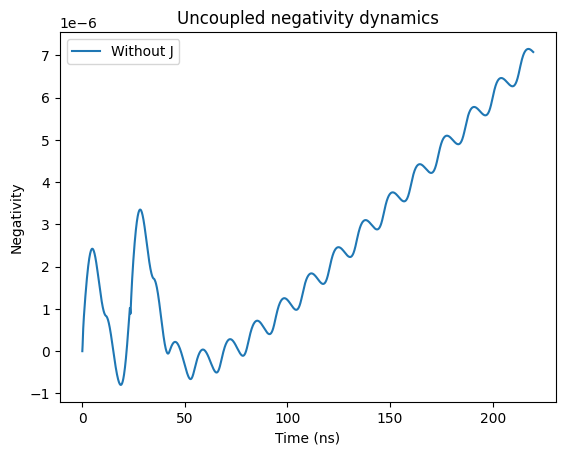}
    \caption{Time evolution of the negativity between the two qubits after deactivating the coupling term $J=0$, while keeping the control pulse active.}
    \label{fig: nega sin J }
\end{figure}

The negativity remains at values on the order of $10^{-6}$, indicating that the state stays essentially separable and that the control pulses, in the absence of coupling, do not generate appreciable entanglement between the qubits. However, small residual oscillations are observed in the signal, with amplitudes that increase slightly over time. These oscillations are consistent with the coherent dynamics associated with the local terms of the Hamiltonian and the frequency difference between the qubits, characterized by the detuning ($\Delta_{78}=v_8-v_7$). Nevertheless, given that the negativity remains virtually zero, these oscillations do not represent significant entanglement generation. The result confirms that, under the conditions considered, the coupling term ($J$) constitutes the necessary mechanism for generating entanglement between the qubits.

\section{Interpretation of the dynamics of negativity in the preparation of the GHZ state}

\begin{figure}[H]
    \centering
    \includegraphics[width=1\linewidth]{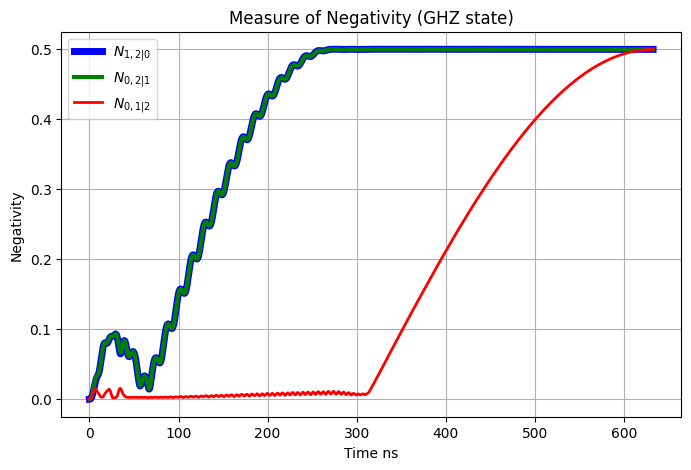}
    \caption{Time evolution of the negativities corresponding to the three bipartitions of the three-qubit system, $N_{12|0}$, $N_{02|1}$, and $N_{01|2}$, during the application of the GHZ square pulse scheme to the closed system (\ref{fig: pulse_ghz_square}). Due to unitary evolution and the absence of decoherence, the curves show a sequential growth of entanglement. During the first stage of the sequence (approximately $0$–$300$ ns), the negativities $N_{12|0}$ and $N_{02|1}$ (blue and green) exhibit early growth, associated with the initial generation of quantum correlations between qubits $0$ and $1$ through channels $D0$, $D1$, and $U0$. Conversely, $N_{01|2}$ (red) remains practically zero during this stage, indicating that the third qubit is not yet significantly involved in multipartite entanglement. Subsequently, the activation of channels $D2$ and $U1$ incorporates the third qubit into the dynamics, causing $N_{01|2}$ to increase and extend the correlations to the three-qubit system. Finally, all three negativities converge to $N=0.5$, the maximum value for bipartitioning one qubit from the other two in an ideal GHZ state, confirming the preparation of the tripartite GHZ state at the end of the control sequence.}
    \label{fig: negativity_GHZ_closed}
\end{figure}

%%%%%%%%%%%%%%%%%%%%%%%%%%%%%%%%%%%%%%%%%

The temporal evolution of the three bipartite negativities reveals the sequential generation of entanglement associated with the two-stage GHZ pulse sequence. During the first stage $t\lesssim 300$ ns, the $D_0$, $D_1$, and $U_0$ pulses, together with the exchange interaction $J_{78}$, generate entanglement between qubits 0 and 1, approximately realizing a state of the form $|\mathrm{Bell}\rangle_{01}\otimes|0\rangle_2$. Consequently, $N_{01|2}$ remains close to zero, while $N_{02|1}$ and $N_{12|0}$ increase and approach $0.5$. The small oscillations observed during this process arise from the coherent interplay between the exchange interaction and the driving fields. Since the dynamics are simulated using the laboratory-frame Hamiltonian with the full carrier-modulated driving fields, without applying the rotating-wave approximation, the observed high-frequency components may also contain contributions from counter-rotating terms retained in the dynamics. In the second stage $t\gtrsim 300$ ns, the $D2$ and $U1$ pulses, associated with $J_{89}$, extend the previously generated entanglement to qubit 2, causing $N_{01|2}$ to increase toward $0.5$. The convergence of all three bipartite negativities to $N\simeq 0.5$ is consistent with the generation of a three-qubit GHZ state, demonstrating that the entanglement is built sequentially, first between qubits 0 and 1 and subsequently extended to the third qubit.
\\\\
\textbf{Code and Numerical Procedure Availability}  The numerical implementation and computational procedures used to simulate quantum dynamics, calculate entanglement measures, and generate the results presented in this work are available in the following \href{https://github.com/delaossaramiro/reparing-pre-established-quantum-correlations-with-pulse-level-control}{GitHub repository} 

\subsection{Differential Evolution (DE)}

Differential Evolution is a numerical optimization method that searches for the best values of a set of parameters by evolving a population of candidate solutions.
The basis idea is;

\begin{itemize}
    \item[1.] Generate candidate solutions : The algorithm starts with several possible solutions, usually generated randomly within predefined bounds.

    \item[2.] Compare and modify solutions:Rather than changing one solution independently, DE uses the differences between existing solutions to determine how to modify them. This allows the population to explore different regions of the parameter space.

    \item[3.] Create new solutions. The modified solution is combined with the original one through a crossover operation, producing a new candidate.

    \item[4.]  Evaluate the candidates: Each candidate solution is evaluated using the objective function to determine its quality.

    \item[5.] Keep the better solution: If the new candidate has a better objective-function value than the current solution, it replaces the current solution. Otherwise, the current solution is retained.

    \item[6.] Repeat: This process is repeated over multiple generations, gradually moving the population toward regions of the search space containing better solutions.

\end{itemize}

The distinctive feature of DE is that the differences between population members provide the search direction. This makes DE a derivative-free method, allowing it to be applied to objective functions for which gradients are unavailable, difficult to calculate, or unreliable.

% \begin{figure}[H]
%     \centering
%     \includegraphics[width= 1\linewidth]{figure/tendencial_lienal-open-2qubits.png}
%     \caption{Caption}
%     \label{fig:placeholder}
% \end{figure}
% \subsection{States obtained for the open two-qubit system}

%%%%%%%%%%%%%%%%%%%%%%%%%%%%%%%%%%%%%%%%%%%%%%%%%%%%%%%%%

% \begin{figure}[t]
%     \centering
%     \includegraphics[width=0.7\linewidth]
%     {distri_kevin.png}
%     \caption{Probability distribution for a $\mathcal{N}_{01} =0.499958 $.}
%     \label{fig 2}
% \end{figure}

% \begin{figure}[t]
%     \centering
%     \includegraphics[width=0.7\linewidth]
%     {esquema_GHZ.png}
%     \caption{Pulse scheme for three qubits.}
%     \label{fig 3}
% \end{figure}

% \begin{figure}[t]
%     \centering
%     \includegraphics[width=0.7\linewidth]
%     {distri_GHZ.png}
%     \caption{Probability distribution for a $\tau = 0.9994767  $.}
%     \label{fig 4}
% \end{figure}

%%%%%%%%%%%%%%%%%%%%%%%%%%%%%%%%%%%%%%%%%%%%%%%%%%%%%%

\newpage

% \begin{figure}[t]
%     \centering
%     \includegraphics[width=0.7\linewidth]
%     {squema_w.png}
%     \caption{Pulse scheme for three qubits.}
%     \label{fig 5}
% \end{figure}

% \begin{figure}[t]
%     \centering
%     \includegraphics[width=0.7\linewidth]
%     {distri_w.png}
%     \caption{Probability distribution for $\mathcal{C}_{01} =0.442009$, $\mathcal{C}_{02} = 0.44433607$ y  $\mathcal{C}_{12} = 0.4421663 $.}
%     \label{fig 6}
% \end{figure}

\newpage

% Bibliografía

%\bibliographystyle{apsrev4-2}
\bibliography{references}

% %TC:endignore

% % Word count
% \verbatiminput{\jobname.wordcount.tex}

% \begin{thebibliography}{9}

% \end{thebibliography}

% %TC:endignore

% % Word count
% \verbatiminput{\jobname.wordcount.tex}

\end{document}